\documentclass[acmsmall,nonacm]{acmart}
\usepackage{acm-ec-26}
\usepackage[ruled]{algorithm2e}
\usepackage{placeins}
\usepackage{bookmark}
\usepackage{tikz}
\usetikzlibrary{arrows.meta}
\setcitestyle{authoryear}
\setcopyright{none}
\AtBeginDocument{%
  \fancypagestyle{firstpagestyle}{%
    \fancyhf{}%
    \fancyfoot[L]{\footnotesize\emph{Working manuscript}}}}

\title{Complete \texorpdfstring{$0.6477$}{0.6477}-EFX Allocations in Strongly Polynomial Time}
\author{Vishwa Prakash HV}
\authornote{\href{mailto:vishwa@keio.jp}{vishwa@keio.jp}}
\affiliation{%
  \institution{Keio University}
  \country{Japan}}
\email{vishwa@keio.jp}

\begin{abstract}
We show that a complete $\alpha$-EFX allocation always exists for
agents with non-negative additive valuations, where
$\alpha \approx 0.647799$ is the unique real root of
$2x^3-2x^2+2x-1=0$. This improves upon the previously known
guarantee of $(\sqrt{5}-1)/2$ for an arbitrary number of agents.
Moreover, we give a strongly polynomial-time algorithm for
computing such an allocation.

Our approach uses a maximum-product matching and an associated
rescaling of the valuations. By constructing three-good bundles
whose two unmatched goods alone guarantee sufficient utility
to their owners, we are able to reassign matched goods and
obtain a partial $\alpha$-EFX allocation in which every agent
values all but at most one unassigned good at no more than
$(1/\alpha-1)$ times the value of their own bundle.
We then develop a generalisation of envy-cycle elimination
that allows decreases in agents' utilities and the return of
previously assigned goods to the unassigned pool, while
preserving $\alpha$-EFX and ensuring termination in a
polynomial number of steps.

The proof was developed using GPT-6.2 Astra in Codex.
The existence theorem has been formally verified using
Lean~4.24.0. The formalisation is available at
\url{https://github.com/severus-tux/alpha_efx_lean}.
\end{abstract}

\IfFileExists{\jobname.aux}{}{
\newlabel{thm:main}{{1.1}{1}{}{theorem.1.1}{}}
\newlabel{cor:rational}{{1.2}{1}{}{theorem.1.2}{}}
\newlabel{sec:related}{{2}{2}{Related work}{section.2}{}}
\newlabel{sec:preliminaries}{{3}{2}{Preliminaries}{section.3}{}}
\newlabel{sec:reduction}{{4}{3}{Reduction, matching and preliminary estimates}{section.4}{}}
\newlabel{lem:positive-core}{{4.1}{3}{Positive matching reduction}{theorem.4.1}{}}
\newlabel{lem:normalise}{{4.2}{4}{Rescaling a maximum-product matching}{theorem.4.2}{}}
\newlabel{eq:prices}{{1}{4}{Rescaling a maximum-product matching}{equation.1}{}}
\newlabel{eq:addition}{{2}{6}{A sufficient condition for adding a good}{equation.2}{}}
\newlabel{lem:cheap}{{4.3}{7}{Trims of a bundle with a low-price good}{theorem.4.3}{}}
\newlabel{sec:overview}{{5}{7}{Proof overview}{section.5}{}}
\newlabel{thm:parametric}{{5.1}{12}{Sufficient numerical conditions}{theorem.5.1}{}}
\newlabel{eq:parameters}{{3}{12}{Sufficient numerical conditions}{equation.3}{}}
\newlabel{eq:row-numbers}{{4}{12}{Sufficient numerical conditions}{equation.4}{}}
\newlabel{eq:watch-parameters}{{5}{12}{Sufficient numerical conditions}{equation.5}{}}
\newlabel{sec:initial}{{6}{13}{The first two allocation passes}{section.6}{}}
\newlabel{lem:two-passes}{{6.1}{13}{Two passes}{theorem.6.1}{}}
\newlabel{eq:first-correlation}{{6}{13}{Two passes}{equation.6}{}}
\newlabel{sec:price-order}{{7}{16}{Exchanges in decreasing order of matched price}{section.7}{}}
\newlabel{lem:initial}{{7.1}{16}{}{theorem.7.1}{}}
\newlabel{eq:initial-correlation}{{7}{16}{}{equation.7}{}}
\newlabel{eq:actual-deletion}{{8}{16}{}{equation.8}{}}
\newlabel{sec:completion}{{8}{20}{Completing the allocation}{section.8}{}}
\newlabel{sec:admissible}{{8.1}{20}{Admissible states}{subsection.8.1}{}}
\newlabel{eq:conditional-trims}{{9}{21}{Admissible states}{equation.9}{}}
\newlabel{lem:complete}{{8.1}{22}{Completion}{theorem.8.1}{}}
\newlabel{eq:outside-source}{{10}{22}{Progress and the graph for an unassigned good}{equation.10}{}}
\newlabel{fig:path}{{1}{25}{Reassignment along a path from $j_0$ to $k$. Each agent before $k$ receives the next agent's old bundle. Agent $k$ receives $S$, which contains $g$ and part of $B_{j_0}$. The goods in $B_{j_0}$ omitted from $S$ return to the pool}{figure.caption.2}{}}
\newlabel{sec:main-proof}{{9}{27}{Proof of the main theorem}{section.9}{}}
\newlabel{sec:parametric-proof}{{9.1}{27}{The parametric guarantee}{subsection.9.1}{}}
\newlabel{sec:parameters}{{9.2}{27}{Choice of the approximation factor}{subsection.9.2}{}}
\newlabel{sec:algorithm}{{10}{29}{The allocation algorithm}{section.10}{}}
\newlabel{sec:algorithm-overview}{{10.1}{29}{The complete algorithm and the computation of prices}{subsection.10.1}{}}
\newlabel{alg:allocation}{{1}{29}{The complete algorithm and the computation of prices}{algocf.1}{}}
\newlabel{alg:initial-passes}{{2}{30}{Constructing the initial partial allocation}{algocf.2}{}}
\newlabel{alg:price-pass}{{3}{31}{Constructing the initial partial allocation}{algocf.3}{}}
\newlabel{alg:completion}{{4}{32}{Trimming a replacement bundle}{algocf.4}{}}
\newlabel{sec:runtime}{{11}{32}{Runtime analysis}{section.11}{}}
\newlabel{thm:runtime}{{11.1}{32}{Running time of \textsc {Allocate}}{theorem.11.1}{}}
\bibcite{afshinmehr2026multi}{{1}{2026}{{Afshinmehr et~al\mbox  {.}}}{{}}}
\bibcite{alkassar2026nine}{{2}{2026}{{Alkassar et~al\mbox  {.}}}{{}}}
\bibcite{amanatidis2024frontier}{{3}{2024}{{Amanatidis et~al\mbox  {.}}}{{}}}
\bibcite{amanatidis2020birds}{{4}{2020}{{Amanatidis et~al\mbox  {.}}}{{}}}
\bibcite{bellman1958routing}{{5}{1958}{{Bellman}}{{}}}
\bibcite{chaudhury2024three}{{6}{2024}{{Chaudhury et~al\mbox  {.}}}{{}}}
\bibcite{chaudhury2020charity}{{7}{2020}{{Chaudhury et~al\mbox  {.}}}{{}}}
\bibcite{farhadi2021envyrank}{{8}{2021}{{Farhadi et~al\mbox  {.}}}{{}}}
\bibcite{filos2026k}{{9}{2026}{{Filos-Ratsikas et~al\mbox  {.}}}{{}}}
\bibcite{hopcroft1973matching}{{10}{1973}{{Hopcroft and Karp}}{{}}}
\bibcite{hv2025types}{{11}{2025}{{HV et~al\mbox  {.}}}{{}}}
\bibcite{hv2026heterogeneity}{{12}{2026}{{HV et~al\mbox  {.}}}{{}}}
\bibcite{lipton2004approximately}{{13}{2004}{{Lipton et~al\mbox  {.}}}{{}}}
\bibcite{plaut2020almost}{{14}{2020}{{Plaut and Roughgarden}}{{}}}
\bibcite{tarjan1972depth}{{15}{1972}{{Tarjan}}{{}}}
\newlabel{tocindent-1}{0pt}
\newlabel{tocindent0}{0pt}
\newlabel{tocindent1}{9.29999pt}
\newlabel{tocindent2}{16.14998pt}
\newlabel{tocindent3}{0pt}
\newlabel{tocindent4}{0pt}
\newlabel{tocindent5}{0pt}
\newlabel{TotPages}{{36}{35}{}{page.35}{}}
\makeatletter
\gdef \@abspage@last{36}
\makeatother
}
\begin{document}
\begin{titlepage}
\maketitle
\end{titlepage}

\section{Introduction}
Envy-freeness up to any good (EFX) is one of the central fairness
notions in the allocation of indivisible goods. An allocation is EFX
if no agent envies another agent's bundle after the removal of any
good from that bundle. Whether a complete EFX allocation, that is, an
EFX allocation in which every good is assigned, always exists for
agents with non-negative additive valuations remains an open question.
This has motivated the study of approximate EFX allocations. For
$0<\alpha\leq 1$, an allocation is $\alpha$-EFX if every agent values
their own bundle at least $\alpha$ times as much as any other bundle
after the removal of any one good. We require this inequality even
when the removed good has value zero to the comparing agent.

Throughout the paper, we write $n$ for the number of agents and
$m$ for the number of goods, and assume that $m>n$, since otherwise
assigning each good to a distinct agent gives a complete EFX
allocation.

For agents with additive valuations, \citet{amanatidis2020birds} and
\citet{farhadi2021envyrank}
established the existence of complete $\alpha$-EFX allocations for
$\alpha=(\sqrt{5}-1)/2\approx 0.618$. We improve this general
approximation guarantee to approximately $0.648$, without imposing
any restriction on the number of agents or the number of distinct
valuation functions.

\begin{theorem}\label{thm:main}
Let $N$ be a finite nonempty set of agents with non-negative additive
valuations on a finite set $M$ of indivisible goods. There exists a
complete $\alpha$-EFX allocation, where $\alpha$ is the unique real root of
\[
 2\alpha^3-2\alpha^2+2\alpha-1=0,\qquad
 \alpha=0.647798871261\ldots\,.
\]
\end{theorem}

For rational valuations, we give a strongly polynomial-time
algorithm for computing such an allocation. More precisely,
Theorem~\ref{thm:runtime} establishes a bound of $O(n^3m^2)$ arithmetic
operations and comparisons. For arbitrary real valuations, the
same operation bound holds in the exact-arithmetic model.

The exact factor also gives a simple rational guarantee.
\begin{corollary}\label{cor:rational}
Every instance of Theorem~\ref{thm:main} admits a complete
$(20/31)$-EFX allocation.
\end{corollary}

Our approach builds on envy-cycle elimination, a standard technique
for extending a partial allocation. If there is a directed cycle of
agents, each of whom envies the bundle of the next agent in the cycle,
we can give each agent the bundle they envy. Such a reassignment
preserves approximate EFX: the collection of bundles is unchanged and
no agent's utility decreases. Repeatedly eliminating envy cycles
allows us to find an agent whose bundle is not envied by any other
agent. Adding an unassigned good to this bundle, however, need not
preserve approximate EFX. For this reason, the partial allocation
from which we start must satisfy additional properties concerning the
values of the unassigned goods.

We first reduce to an instance in which every agent can be matched
to a positively valued good. We then compute a maximum-product
matching and rescale each agent's valuation independently. After
rescaling, every agent values their matched good at least one,
every unmatched good is worth at most one to every agent, and the
value of a matched good to its original matched agent bounds its
value to all other agents. These bounds reveal the difficulty in
improving the golden-ratio approximation:
some matched
goods are too valuable to be placed in a bundle of two goods without
violating approximate EFX for agents with low utility, yet their
original owners do not have enough utility for the usual sufficient
condition for assigning the remaining goods to hold. Section~\ref{sec:overview}
explains the precise interval of matched-good values for which this
difficulty arises.

Our first contribution is a construction that deals with this
interval by reassigning matched goods. We form three-good bundles
whose two unmatched goods together give their owner enough utility to
satisfy the approximate EFX comparisons, independently of the value
of the matched good. This enables us to replace the matched good
later. We show how to use these matched goods to compensate other
agents and establish a partial allocation in which at most one
unassigned good per agent violates the usual sufficient condition
for insertion. For each agent with such an exception, we also obtain
bounds on the values of all the other unassigned goods and on the
value of every allocated bundle after removing one good.

Our second contribution is a completion procedure that uses these
bounds to allow decreases in some agents' utilities and the return
of previously assigned goods to the unassigned pool. This is a
departure from ordinary envy-cycle elimination, where utilities
are non-decreasing and goods, once assigned, remain assigned.
We show that a suitable lexicographic measure of progress strictly
improves under every operation, even if an individual operation
decreases utility or releases goods. A further analysis of how
agents' rankings of the current bundles change under these
operations gives a strongly polynomial bound on the number of
iterations.

\section{Related work}\label{sec:related}

\citet{plaut2020almost} gave a complete $1/2$-EFX guarantee for
subadditive valuations, under which a union is worth at most the
sum of the values of its two sets. This class includes non-negative
additive valuations.
\citet{amanatidis2020birds} improved the guarantee for additive
valuations to $\varphi-1=(\sqrt5-1)/2$, where
$\varphi=(1+\sqrt5)/2$. Their method extends a draft allocation
by rotating bundles among agents who each prefer the next agent's
bundle to their own. \citet{lipton2004approximately} introduced this
technique for envy-freeness up to one good (EF1).
\citet{farhadi2021envyrank} obtained the same EFX factor using a
matching that maximises the product of the values of the matched
goods to their owners. We use this matching and the associated rescaling of valuations to
construct our initial partial allocation. Our completion procedure
generalises envy-cycle elimination by permitting reassignments that
decrease some agents' utilities and release previously allocated goods,
while preserving the required EFX inequalities.

\citet{amanatidis2024frontier} obtained complete $2/3$-EFX for up to
seven agents, and also for several restricted classes of additive
valuations. A valuation type is a set of agents with the same
valuation function. \citet{hv2026heterogeneity} obtained $2/3$-EFX for
at most four types, with arbitrary multiplicities. More recently,
\citet[Theorem~3.7]{filos2026k} extended the seven-agent guarantee
to eight agents. The same paper gives a $3/4$ guarantee for
$\mathrm{EF2X}$ with any number of agents, under comparisons that
delete two goods. In contrast to the guarantees that restrict the number of agents or
valuation types, Theorem~\ref{thm:main} applies to arbitrary non-negative
additive valuations and requires the EFX comparisons after the removal
of any single good.

\citet{chaudhury2024three} proved exact EFX for three agents with
additive valuations. \citet{hv2025types} proved exact EFX for three
valuation types, with arbitrary multiplicities. \citet{alkassar2026nine}
established exact EFX for four additive agents and at most nine
goods. In the multigraph setting, goods are edges. The only agents who
can value a good positively are the two endpoints of its edge.
\citet{afshinmehr2026multi} proved exact EFX for multigraphs with
additive valuations.
Another line of work permits leaving goods unassigned. For example,
\citet{chaudhury2020charity} obtain exact EFX with at most $n-1$
unassigned goods and no agent envying the unassigned set.

\section{Preliminaries}\label{sec:preliminaries}

Let $N$ be the set of agents and $M$ the set of goods. For each
agent $i\in N$, let $v_i(g)\geq 0$ denote the value of good
$g\in M$ to $i$. By additivity,
\[
    v_i(S)=\sum_{g\in S}v_i(g)
    \qquad(S\subseteq M).
\]
In particular, $v_i(\varnothing)=0$, and $v_i(S)\leq v_i(T)$
whenever $S\subseteq T$.

A \emph{partial allocation} $B=(B_i)_{i\in N}$ is a family of
pairwise disjoint bundles $B_i\subseteq M$. A bundle may be empty,
and we call a bundle with exactly one good a singleton. The
\emph{pool} $P(B)=M\setminus\bigcup_{i\in N}B_i$ consists of the
goods currently unassigned. An allocation is \emph{complete} if
$P(B)=\varnothing$.

\begin{definition}[Approximate EFX]
Let $0<\alpha\le1$. A partial allocation $B$ is $\alpha$-EFX if
\[
  \forall i,j\in N\;\;\forall h\in B_j,\qquad
  v_i(B_i)\ge\alpha\,v_i(B_j\setminus\{h\}).
\]
At $\alpha=1$ we call the allocation EFX. For $i=j$ the inequality
follows from non-negativity and $\alpha\le1$.
\end{definition}

We will frequently consider bundles obtained by removing one good.
For a bundle $T$ and a good $h\in T$, we call $T\setminus\{h\}$ a
\emph{trim} of $T$. An empty bundle has no trims, and a singleton has
only the empty set as a trim. Thus an allocation is $\alpha$-EFX if
every agent values their own bundle at least $\alpha$ times as much as
every trim of every other bundle. Notice that the definition includes
removing a good of value zero to the comparing agent. By additivity,
for a fixed agent and a nonempty bundle, the most valuable trim is
obtained by removing one of the least valuable goods of that bundle.
In comparison, envy-freeness up to one good (EF1) only requires the
EFX inequality to hold for at least one choice of the removed good
from each nonempty bundle, at $\alpha=1$.

\section{Reduction, matching and preliminary estimates}\label{sec:reduction}

\subsection{A reduction to a positive matching}

Our matching construction requires that each agent can be assigned
a distinct good they value positively. This condition may fail
when some singleton values are zero. We first show that such
instances can be reduced to instances satisfying the condition,
without losing any of the EFX guarantees on the original instance.

Consider the bipartite graph with vertex sets $N$ and $M$, and an
edge $(i,g)$ whenever $v_i(g)>0$. Let $F$ be a maximum-cardinality
matching in this graph. If $F$ does not match every agent, we will
set aside a group of agents and goods as follows. Orient the edges
of $F$ from goods to agents, and every other edge from agents to
goods. Starting from all unmatched agents, let $L\subseteq N$ and
$T\subseteq M$ be the sets of agents and goods reachable in this
directed graph. If there is no unmatched agent, both sets are empty.
Write
\[
    I=N\setminus L,\qquad G=M\setminus T.
\]
For each matched agent in $L$, assign them their good in $F$ as a
singleton. Leave the unmatched agents of $L$ with empty bundles
for now.

\begin{lemma}[Positive matching reduction]\label{lem:positive-core}
The restriction of $F$ to $I$ matches every agent in $I$ to a
positive-valued good in $G$. Moreover, $|G|>|I|$, and every agent
in $L$ values every good of $G$ at zero.

If $I\ne\varnothing$, any complete $\alpha$-EFX allocation on the
residual instance $(I,G)$, together with the bundles assigned to
$L$, is a complete $\alpha$-EFX allocation of the original
instance. If $I=\varnothing$, assigning all the goods of $G$ to
any unmatched agent of $L$ completes the allocation and gives EFX.
\end{lemma}

\begin{proof}
No good in $T$ can be unmatched in $F$, since a shortest directed
path from an unmatched agent to an unmatched good would be an
augmenting path, contradicting the maximal cardinality of $F$.
The matched agent of every good in $T$ is also reachable, and hence
belongs to $L$.

Every positive-valued neighbour of an agent in $L$ belongs to $T$.
For an edge not in $F$, this follows from its orientation from
agents to goods. If the edge belongs to $F$, its agent must have
been reached through that edge from its matched good. Conversely,
every reached good is reached from a positive-valued neighbour in
$L$. Thus $T$ is exactly the positive neighbourhood of $L$.
Consequently,
\[
    v_i(g)=0\qquad(i\in L,\ g\in G).
\]
The matching $F$ pairs the goods in $T$ bijectively with the
matched agents in $L$. In particular, $|T|\leq |L|$. Every
unmatched agent is in $L$, so every agent in $I$ is matched in $F$
to a good in $G$. Furthermore,
\[
    |G|-|I|=m-n+|L|-|T|>0.
\]
Here we use the standing assumption $m>n$.

Suppose first that $I\ne\varnothing$. In the resulting allocation,
an agent in $L$ values every bundle assigned to an agent in $I$
at zero. Every bundle assigned to $L$ is empty or a singleton,
so its trims, if any, are empty. Hence all EFX comparisons involving
an agent of $L$ hold, regardless of the allocation on $I$. The
remaining comparisons hold by the assumed $\alpha$-EFX property on
$(I,G)$. Every good in $T$ is assigned by $F$, and every good in $G$
is assigned by the residual allocation. The combined allocation
is therefore complete and $\alpha$-EFX.

If $I=\varnothing$, then $L=N$. Since $N$ is nonempty and the
search reaches every agent, it must have started from at least
one unmatched agent. Every good in $G$ is now worth zero to every
agent. Assign all these goods to one of the unmatched agents in
$L$. All other bundles are empty or singletons, and every trim
of the bundle containing $G$ is worth zero to every agent. The
resulting allocation is complete and EFX.
\end{proof}

In the remaining construction we work with the residual instance
when $I\ne\varnothing$. To simplify notation, we again write $N$ and
$M$ for its sets of agents and goods, and $n$ and $m$ for their
cardinalities. The preceding lemma guarantees that the standing
assumption continues to hold and that there is a matching which
assigns every agent a positively valued good.

\subsection{Matching and rescaling}

We now construct an initial allocation in which every agent
receives one good. The particular matching and a suitable
rescaling of the valuations will give us useful bounds on the
values of both matched and unmatched goods.

For each agent $i$, multiplying their valuation $v_i$ by a positive
constant $\lambda_i$ preserves all the EFX inequalities for that
agent. Thus we can choose these scaling factors independently,
without changing which allocations are $\alpha$-EFX. We use the
following matching argument, based on the normalisation of
\citet{farhadi2021envyrank}.

\begin{lemma}[Rescaling a maximum-product matching]\label{lem:normalise}
Suppose that there exists an injective map $c:N\to M$ such that
$v_i(c(i))>0$ for every agent $i\in N$.

Let $(a_i)_{i\in N}$ be a maximum-product matching between
agents and goods; that is, the goods $a_i$ are distinct and the
matching maximises
\[
    \prod_{i\in N}v_i(a_i)
\]
over all matchings that assign one good to each agent. Write
$U=M\setminus\{a_i:i\in N\}$ for the unmatched goods.

Then there exist positive scaling factors $(\lambda_i)_{i\in N}$
such that, for the rescaled valuations $w_i=\lambda_i v_i$ and the
prices $\pi_i=w_i(a_i)$, the following inequalities hold:
\begin{equation}\label{eq:prices}
\begin{aligned}
    \pi_i &\geq 1
        &&\text{for every }i\in N,\\
    w_i(a_j)&\leq \pi_j
        &&\text{for every }i,j\in N,\\
    w_i(g)&\leq 1
        &&\text{for every }i\in N\text{ and }g\in U.
\end{aligned}
\end{equation}
\end{lemma}

The first inequality ensures that every agent has utility at
least one when they hold their matched good. The second associates
each matched good with a fixed upper bound on its value to every
agent, equal to its value to its original matched agent. The third
bounds the value of every unmatched good by one, for every agent.
These properties allow us to bound the value of a bundle in terms
of the prices of its matched goods and the number of unmatched
goods it contains.

Throughout the proof, $a_i$ denotes the good originally matched
to agent $i$, even if it is later assigned to another agent. Its
price $\pi_i$ remains fixed. Similarly, a good outside the matching
is called unmatched even after it has been allocated.

\begin{proof}
The assumed positive matching guarantees that the maximum product
is positive, so $v_i(a_i)>0$ for every $i\in N$.

Introduce $m-n$ dummy agents, each of whom values every good at
one. Extend the fixed matching $(a_i)_{i\in N}$ to a bijection
between the enlarged set of agents and the goods, by matching
the dummy agents to the goods of $U$. This is a maximum-product
bijection: the dummy agents contribute a factor of one to the
product, and the matching of the real agents is already optimal.
For the remainder of this proof, write $a_i$ for the good matched
to agent $i$ in the enlarged instance, including the dummy agents.

Consider a directed graph on the enlarged set of agents. Whenever
$v_i(a_j)>0$, introduce an edge $i\to j$ of weight
$v_i(a_j)/v_i(a_i)$. The weight is well-defined because every
matched value is positive. The weight of a directed walk is the
product of its edge weights, with the empty product equal to one.

Every directed simple cycle has weight at most one. Indeed,
reassigning the matched goods along the cycle multiplies the
product of the matching by precisely the weight of that cycle.
A cycle of weight greater than one would therefore contradict
the optimality of the matching.

If a walk repeats a vertex, consider its first repeated vertex.
The closed portion between its two occurrences is a simple cycle.
Deleting this cycle preserves the terminal vertex and cannot
decrease the weight of the walk, because the removed cycle has
positive weight at most one. Repeating this operation
produces a walk with no repeated vertices. Consequently, the
maximum weight $p_j$ of a directed walk ending at $j$, with no
restriction on its starting vertex, exists: it is attained among
the finitely many walks with no repeated vertices.

The walk of length zero gives $p_j\geq 1$ for every $j$. Appending
an edge $i\to j$ to a walk attaining $p_i$ gives
\[
    p_i\frac{v_i(a_j)}{v_i(a_i)}\leq p_j.
\]
The same inequality holds when $v_i(a_j)=0$, because its left-hand
side is then zero.

Every edge out of a dummy agent has weight one. Hence, if $i$ is
a dummy agent, the preceding inequality gives $p_i\leq p_j$ for
every $j$ in the enlarged instance. In particular, all the dummy
agents share a common minimum walk weight, which we denote by
$p^*$. There is at least one dummy agent by the standing
assumption.

For each real agent $i$, set
\[
    \lambda_i=\frac{p_i}{p^*v_i(a_i)}.
\]
These factors are positive. We have
\[
    w_i(a_i)=\frac{p_i}{p^*}\geq 1.
\]
For every real agent $i$ and every good $a_j$ in the enlarged
matching, the walk inequality gives
\[
    w_i(a_j)
    =\frac{p_i}{p^*}\frac{v_i(a_j)}{v_i(a_i)}
    \leq\frac{p_j}{p^*}.
\]
If $j$ is a real agent, the right-hand side is $w_j(a_j)=\pi_j$.
If $j$ is a dummy agent, the right-hand side is one. Discarding the
dummy agents leaves exactly the original maximum-product matching
and establishes all three inequalities.
\end{proof}

\subsection{A sufficient condition for adding a good}

Throughout the proof, write $r=1/\alpha$ for the reciprocal of
the approximation factor. We consider $3/2\leq r\leq 2$. For a
partial allocation $B$, write $u_i=w_i(B_i)$ for the utility of
agent $i$ with respect to the rescaled valuation. To establish
$(1/r)$-EFX, it suffices to show that every trim of every bundle
has value at most $ru_i$ to agent $i$.

Suppose we wish to add an unassigned good $g$ to a bundle $T$
that agent $i$ does not envy. If $w_i(g)\leq(r-1)u_i$, then
\[
    w_i(T\cup\{g\})
    =w_i(T)+w_i(g)
    \leq u_i+(r-1)u_i
    =ru_i.
\]
Since the valuations are non-negative, every trim of the enlarged
bundle is also worth at most $ru_i$ to $i$. This is a sufficient
condition for adding a good to an unenvied bundle while preserving
the EFX comparisons of agent $i$.

By Lemma~\ref{lem:normalise}, every unmatched good is worth at most one
to every agent. Hence an agent whose utility is at least
\[
    q=\frac{1}{r-1}
\]
satisfies the preceding condition for every unmatched good.

Consider now an agent $j$ who holds their original matched
singleton and whose matched price is less than $q$. If we add an
unmatched good to this singleton, removing the added good leaves
$\{a_j\}$ as a trim. Since $w_i(a_j)\leq\pi_j<q$ for every agent
$i$, an observer of utility at least
\[
    b=\frac{q}{r}
\]
can satisfy the EFX comparison with this trim. Thus the agents
with utility below $b$ are the only ones who can obstruct such a
pairing. Our initial construction will increase the utility of agents who
receive certain new bundles to at least $b$, while preserving the
original matched singleton of every agent whose utility remains
below $b$. The numerical conditions used in the proof ensure
$1\leq b\leq r$.

During completion we will allow a transfer to an agent only if
it gives them enough utility both to preserve their existing
insertion conditions and, when necessary, to reach $b$. Therefore,
when we choose a bundle to which we wish to add a good, we will
sometimes have an upper bound of $\max\{b,u_i\}$ on its value to
an observer, rather than just $u_i$. We accordingly strengthen
the sufficient condition above.

We say that a good $g$ satisfies the \emph{insertion condition}
for agent $i$ at utility $u_i$ if
\begin{equation}\label{eq:addition}
    \max\{b,u_i\}+w_i(g)\leq ru_i.
\end{equation}
If $w_i(T)\leq\max\{b,u_i\}$, this condition ensures that the
whole of $T\cup\{g\}$, and hence each of its trims, is worth at
most $ru_i$ to $i$.

Equivalently, the insertion condition consists of the inequalities
\[
    b+w_i(g)\leq ru_i,
    \qquad
    w_i(g)\leq (r-1)u_i.
\]
Both remain satisfied if the agent's utility increases. When
$u_i\geq b$, the second inequality implies the first.

\subsection{Trims of a bundle with a low-price matched good}

In the initial construction we will form bundles consisting of a
matched good of price below $b$ and at most two unmatched goods. The
same kind of bundle may arise during completion. Every trim of such a
bundle contains either at most two unmatched goods, or the matched
good and at most one unmatched good. This gives the uniform upper
bound $b+1$ on the value of a trim to any agent. An agent whose utility
is at least $(b+1)/r$ can therefore satisfy every EFX comparison with
such a bundle. For agents with smaller utilities, the following lemma
provides a second bound, using their insertion conditions on the
unmatched goods.

\begin{lemma}[Trims of a bundle with a low-price good]\label{lem:cheap}
Suppose the rescaled valuations satisfy \eqref{eq:prices}. Let $T$
consist of one matched good of price below $b$ and at most two
unmatched goods. Then
\[
  \forall i\in N\;\;\forall h\in T,\qquad
  w_i(T\setminus\{h\})\le b+1.
\]
Fix an agent $i$ and a utility $u_i\ge1$. If every unmatched good of
$T$ satisfies the insertion condition~\eqref{eq:addition} for $i$ at $u_i$, then
$w_i(T\setminus\{h\})\le ru_i$ for every $h\in T$.
\end{lemma}

\begin{proof}
The bound \eqref{eq:prices} gives
$\max\{2,b+1\}=b+1$ as an upper bound on the value of every trim.
If we delete the matched good and the unmatched goods satisfy
\eqref{eq:addition}, their total value is at most
$2(r-1)u_i\le ru_i$, since $r\le2$. If the matched good and one unmatched
good $g$ remain, their combined value is at most $b+w_i(g)\le ru_i$.
A matched good alone has value below $b\le r\le ru_i$.
\end{proof}

\subsection{Directed graphs and bundle transfers}

During completion, we will assign each agent $i$ a threshold $t_i$ for
the value of a bundle they can receive. We draw a directed edge
$i\to j$ between distinct agents if $w_i(B_j)>t_i$. The thresholds will
be chosen so that a transfer along an edge helps us maintain the
properties of our partial allocation. The absence of an edge is also
useful: it bounds the value of a bundle to an agent who does not
participate in a transfer.

Recall that a \emph{source component} of a directed graph is a strongly
connected component with no incoming edge from outside the component.
Every finite nonempty directed graph has a source component, since the
graph obtained by contracting its strongly connected components is
acyclic. For the graph just described, if $C$ is a source component,
then
\[
    w_i(B_j)\leq t_i\qquad(i\notin C,\ j\in C).
\]
This lets us bound the values of bundles in $C$ to agents outside $C$.
Within $C$, we can use directed paths to transfer bundles among its
members. A component may have more than one vertex even when the graph
has no vertex of indegree zero. If it has at least two vertices, it
contains a directed cycle, and every vertex in it has an incoming edge
from another member.

We will use two kinds of transfers. Along a directed cycle, give every
agent the old bundle of their successor. This changes the owners of
bundles but not the bundles themselves. Along a simple directed path
$j_0,j_1,\ldots,j_\ell$, give $j_t$ the old bundle $B_{j_{t+1}}$ for each
$0\leq t<\ell$ and give $j_\ell$ a replacement bundle $S$. This is
also well-defined for a path of length zero, when only the bundle of
$j_0$ is replaced. If $S$ consists of some of the goods of $B_{j_0}$
and one currently unassigned good, then the resulting bundles are
pairwise disjoint: the only old bundle not transferred whole is
$B_{j_0}$. The goods of $B_{j_0}$ omitted from $S$ return to the pool.

\section{Proof overview}\label{sec:overview}

The proof has two parts. We first construct a partial
$(1/r)$-EFX allocation in which, for each agent, at most one
unassigned good violates the insertion condition. For agents with
such a good, we then strengthen the bounds on the values of the
other unassigned goods and of all existing trims. The main idea
is to reassign the matched goods in certain three-good bundles
while keeping their owners' utilities above a suitable threshold.

The second part uses these bounds to complete the allocation.
Some of the required reassignments may decrease utilities or
return previously allocated goods to the unassigned pool. We
explain why approximate EFX can nevertheless be preserved, and
why these operations lead to a complete allocation. Finally, we
derive the value of $r$ from the inequalities used in the two parts
of the proof.

\subsection{The gap between two price ranges}

Start with the allocation in which each agent holds their original
matched good. By Lemma~\ref{lem:normalise}, every agent has utility at least one,
and every unmatched good is worth at most one to every agent. There
are two price ranges in which we can handle the unassigned goods
without much difficulty.

If the price of a matched good $a_i$ is at most $r$, we can add an
unmatched good to $\{a_i\}$. Each trim of the resulting pair consists
of one good worth at most $r$ to every agent. Since all utilities are
at least one, forming such a pair cannot violate the EFX comparisons
with it. Giving an agent their favourite unassigned good in this way
can raise their utility enough that all the remaining goods satisfy
their insertion condition.

At the other end of the range, an agent whose original matched good
has price at least $q=1/(r-1)$ already satisfies the insertion condition
for every unmatched good: every such good is worth at most one, and
$1\leq(r-1)q$. From this agent's perspective, we do not need to increase
their utility before assigning the remaining goods to suitably valued
bundles.

The two ranges cover all matched prices if and only if
\[
    r\geq\frac{1}{r-1},\qquad\text{or equivalently}\qquad
    r(r-1)\geq1.
\]
This explains the golden-ratio approximation in the earlier matching
and completion approach. At equality, $1/r=(\sqrt{5}-1)/2$. If we seek
a stronger approximation, then $r<q$, leaving the interval $(r,q)$.
The difficulty is that an agent in this interval need not satisfy the
insertion condition for their favourite unassigned good, but simply
adding that good to their singleton is not necessarily safe either.

To understand the obstruction, consider an agent $i$ whose matched
price $\pi_i$ lies in $(r,q)$ and an unassigned good $g$. Forming the
pair $\{a_i,g\}$ would expose $\{a_i\}$ as a trim. Suppose this trim
violates the EFX inequality for an agent $k$. Then
\[
    ru_k<w_k(a_i)\leq\pi_i<q,
\]
so $u_k<\pi_i/r<b=q/r$. We arrange the initial allocation so that any
agent whose utility is below $b$ still holds their original matched
good alone. Hence $B_k=\{a_k\}$, and the price of $a_k$ is below $b$.
We can give $a_i$ to agent $k$ as a singleton, eliminating the
obstruction: a singleton has only the empty trim. The question is
how to compensate agent $i$, who may value $a_k$ very little.

Suppose that two unassigned goods $g,h$ are sufficiently valuable to
$i$. We can exchange the matched goods and give $i$ both of these
unassigned goods, resulting in
\[
    B_k'=\{a_i\},\qquad B_i'=\{a_k,g,h\}.
\]
Since the price of $a_k$ is below $b$, every trim of the new
three-good bundle has value at most $\max\{2,b+1\}=b+1$ to every
observer. The same upper bound holds for all pairs formed in the
initial allocation, since their trims are worth at most $q\leq b+1$.
Consequently, an agent whose utility is at least
\[
    d_0=\frac{b+1}{r}
\]
can satisfy every EFX comparison at this stage. For observers with
smaller utilities, we use the insertion conditions obtained by
assigning favourite goods to agents in the low-price range.

We require the two unmatched goods in the triple to give its owner
utility at least $d_0$, independently of the value of $a_k$. There are
two reasons for this requirement. First, the recipient has enough
utility to compare with all the initial trims even if they value the
new matched good very little. Second, we may later replace $a_k$ with
another low-price matched good without reducing the recipient's
utility below $d_0$. This second property is crucial to the next part
of the construction.

If an agent's favourite unassigned good violates their insertion
condition but their two favourite unassigned goods together are worth
less than $d_0$, we can deduce something useful even without assigning
either good. If the favourite is worth $z$, every other unassigned
good is worth less than $d_0-z$ to this agent. Our parameter choices
ensure that all those other goods satisfy the insertion condition.
Thus, after considering each agent, there is at most one unassigned
good per agent for which the insertion condition can fail. We call
this the agent's \emph{exceptional good}. The formal construction of
Lemma~\ref{lem:two-passes} uses a slightly more conservative rule for choosing $k$
in the exchange, namely that $k$ holds their original singleton of
price below $b$ and values $a_i$ at least $b$. This covers every actual
obstruction and ensures that the recipient reaches utility at least
$b$. In the next stage we use the actual EFX obstruction to obtain a
sharper estimate on the price of the singleton we receive in exchange.

\subsection{Reusing the matched good of a second-pass triple}

The preceding construction gives a useful partial allocation, but it
may still leave agents who have an exceptional good. We want to
strengthen the bounds for these agents before attempting to complete
the allocation. In particular, the uniform threshold $d_0$ is more
than we need for some of them. If we can use a lower utility threshold in the exchange, then
failing to find a sufficiently valuable pair of unassigned goods
implies a stronger upper bound on the value of every good other
than the favourite. This is the bound we seek before completion.

Let $R_0$ be the set of goods left unassigned by the first two passes.
Consider an agent $i$ who still holds their original singleton
$\{a_i\}$ and whose exceptional good $g_i\in R_0$ is worth
$z_i=w_i(g_i)$. Write $p=\pi_i\in(r,q)$. If giving $g_i$ to $i$ would
violate another agent $k$'s EFX inequality, we know from the preceding
argument that $u_k<p/r$. Consequently, the matched good $a_k$ that we
obtain by giving $a_i$ to $k$ has price below $p/r$, not just below $b$.
A triple containing $a_k$ and two unmatched goods therefore has all
its trims worth at most $1+p/r$ to every observer. This suggests that
agent $i$ need only receive utility
\[
    D_i=D(p)=\frac{1+p/r}{r},
\]
provided they can afford all the other bundles at this level of
utility. Since $p<q$ and $b=q/r$, we have $D_i<d_0$.

What if some existing trim is worth more than $rD_i=1+p/r$ to $i$?
This is precisely where the earlier three-good bundles become useful.
Because $p>r$, the value of such a trim exceeds two. It cannot come
from a singleton, or from a pair whose goods are each worth at most
$q<2$. Nor can it consist of two unmatched goods, which together are
worth at most two. It cannot come from a triple created in the
current stage either: both unmatched goods of such a triple belong
to $R_0\setminus\{g_i\}$ while $g_i$ is still unassigned. By the
previous construction, each of these goods is worth less than
\[
    s_0=d_0-(r-1)r
\]
to $i$, and our parameter choices ensure
$\max\{b+s_0,2s_0\}\leq2$.

We conclude that the trim must come from a triple constructed in the
second pass, and it must contain that triple's current matched good,
say $\ell$. The other good in the trim is unmatched and worth at most
one. Therefore $w_i(\ell)>p/r>1$. Importantly, we can give $\ell$ to
$i$ without depriving the owner of the triple of the utility needed
for approximate EFX, because that owner retains the two original
unmatched goods that alone give them utility at least $d_0$.

Suppose the direct pair $\{a_i,g_i\}$ is unsafe, and let $k$ be an
agent for whom the pair would violate an EFX inequality. Let $j$ own
the second-pass triple $\{\ell,e,f\}$ identified above. We can perform
a three-agent exchange: give $a_i$ alone to $k$, give $\ell$ together
with $g_i$ to $i$, and replace $\ell$ with $a_k$ in $j$'s triple.
Agent $k$ strictly improves their utility to above $r$, agent $i$
receives utility greater than $1+(r-1)r\geq d_0$, and agent $j$ retains
both $e$ and $f$ and hence utility at least $d_0$. The only new good
assigned is $g_i$. The crucial feature of this operation is that we
replace the matched good $\ell$ in an existing triple by a matched good
whose price is below $\pi_i/r$.

On the other hand, suppose $i$ already values every old trim at most
$rD_i$. Then there is no need to take a matched good from a triple. If
their two favourite unassigned goods $g_i,h$ together are worth at
least $D_i$, we can perform the simpler exchange in which $k$ receives
$\{a_i\}$ and $i$ receives $\{a_k,g_i,h\}$. Agent $i$ can afford the old
trims at utility $D_i$. Unlike a triple formed in the second pass, this
new triple may not provide its owner with utility $d_0$ independently
of its matched good. We therefore do not replace its matched good
during this stage. In either case, if action is called for but the
direct pair is safe, we simply give $i$ the pair $\{a_i,g_i\}$.

There is one remaining issue: a later exchange could invalidate the
trim bound of an agent considered earlier. We avoid this by considering
agents in nonincreasing order of their original matched prices.
Suppose $v$ is considered before $i$. In an exchange involving $i$, the
matched good inserted into a second-pass triple has price less than
$\pi_i/r\leq\pi_v/r$. Every trim of that modified triple is therefore
worth at most $1+\pi_v/r=rD_v$ to $v$. A triple created later in the
current stage uses two goods of $R_0\setminus\{g_v\}$, provided $g_v$
is either still unassigned or was assigned to $v$ earlier. Its trims
are worth at most two to $v$, and new pairs have trims worth at most
$q<2$. These bounds protect both an earlier agent who accepted a
bundle at utility $D_v$ and an earlier agent whose exceptional good
remains unassigned. The same second-pass triple may be modified
several times; its owner retains the original two unmatched goods
throughout.

We perform an operation for an agent $i$ only if some old trim is
worth more than $rD_i$ to $i$ or if their two favourite unassigned
goods together are worth at least $D_i$. If their exceptional good
is still unassigned when the construction ends, neither condition
held when we considered them. As the pool only shrinks, and the trim
bound is preserved by the decreasing-price order, we obtain the
following two bounds at the end of the construction, where $R$ denotes
the final pool:
\[
    w_i(h)<D_i-z_i\qquad(h\in R\setminus\{g_i\}),
    \qquad
    w_i(B_j\setminus\{h\})\leq 1+\frac{\pi_i}{r}
        \qquad(j\in N,\ h\in B_j).
\]
The first inequality bounds the values of all the unassigned goods
other than $g_i$, and the second bounds how much utility $i$ needs in
order to satisfy their EFX comparisons with the current bundles.
Both bounds are specific to agent $i$. Lemma~\ref{lem:initial} establishes them
formally while preserving approximate EFX.

\subsection{Completing the allocation}

We now explain how the two bounds obtained above allow us to complete
the allocation. Suppose that every trim of every current bundle has
value at most $K_i$ to an agent $i$. If we give $i$ a bundle of value
at least $K_i/r$, all their EFX comparisons with the surviving old
bundles remain satisfied, even if their utility decreases. We need
only establish their comparisons with any newly formed bundle. This
observation allows us to depart from ordinary envy-cycle elimination.

Fix the pool $R$ left by the initial construction, and for each agent
who has an exceptional good at this point, fix that good $g_i$ and its
value $z_i=w_i(g_i)$. Let $s_i=D_i-z_i$. Every other good in $R$ is
worth less than $s_i$ to $i$. We maintain a set $W$ of agents whose
exceptional goods still require special treatment. For each $i\in W$,
we bound the value to $i$ of every trim of every bundle not containing
$g_i$ by $K_i$. Whenever $g_i$ is unassigned, this bound applies to
all the old bundles, so utility $d_i=K_i/r$ suffices for comparisons
with them. We will choose $K_i$ below.

There is an important distinction between assigning an exceptional
good temporarily and ensuring that it remains assigned. Since some
operations return goods to the pool, we cannot check the insertion
condition only for the currently unassigned goods. We distinguish a
set $A$ of agents outside $W$ with utility at least $b$. We will ensure
that no agent leaves $A$, and that the goods in bundles of $A$ can be
transferred with their bundles but can never return to the pool.
For each agent outside $W$, we maintain the insertion condition on
all goods of $R$ not held by members of $A$. In particular, once $g_i$
is held by an agent in $A$, we need only check the insertion condition
for the other potentially releasable goods. If $i$ has utility at
least $d_i\geq b$ and $s_i\leq(r-1)d_i$, these goods satisfy the
condition and $i$ can leave $W$.

Fix an unassigned good $g$ that we wish to allocate. We use a directed
graph whose edges describe possible transfers of bundles. An agent
outside $W$ must receive a bundle worth more than $\max\{b,u_i\}$, so
their utility increases and they can join or remain in $A$. An agent
tracking an exceptional good other than $g$ requires utility above
$q$: at this utility the insertion condition holds even for their
exceptional good. For an agent tracking $g$ itself we can use the
lower threshold $d_i$, provided the operation assigns $g$ permanently
and satisfies the new EFX comparisons.

Draw an edge $i\to j$ if $i$ values $j$'s bundle strictly above their
receiving threshold. We use a source strongly connected component of
this graph because its members can transfer bundles along directed
paths, while every outside agent values every bundle in the component
at most at their receiving threshold.

The source-component property explains the choice of $K_i$. The
initial construction gives the bound $1+\pi_i/r$ on all trims. When
we add a good different from $g_i$ to a bundle in a source component
that does not contain $i$, the old bundle is worth at most $q$ to
$i$, and the added good is worth less than $s_i$. The same quantity
will bound the new bundles constructed inside the component. Thus we
set
\[
    K_i=\max\left\{1+\frac{\pi_i}{r},q+s_i\right\},
    \qquad d_i=\frac{K_i}{r}.
\]
For an outside agent tracking $g$ itself, the enlarged bundle is
instead worth at most $d_i+z_i$. The sufficient condition
$d_i+z_i\leq r\pi_i$ will determine our approximation factor, since
an agent still tracking an exceptional good retains their original
matched good and therefore has utility at least $\pi_i$.

If the source component consists of one agent, we can simply add
$g$ to their bundle. If it has more than one agent but none tracks
$g$, rotating bundles along a directed cycle strictly increases
utility for each participant. The remaining case is the difficult
one: a cycle rotation might decrease the utility of an agent tracking
$g$ without assigning that good permanently.

Instead, choose a starting agent in the component whose bundle can
still release goods. We retain their original matched good, add $g$,
and retain an inclusion-minimal subset of the remaining goods that
makes the replacement acceptable to at least one member of the
component. Because minimality is taken over all possible recipients,
removing any one of the retained goods makes the replacement
unacceptable to every member of the component. These inequalities,
together with the insertion condition and the bounds $s_i$, control
the value of the replacement after any single deletion. We use a
directed path from the starting agent to an accepting recipient to
transfer the other bundles whole and give the replacement to the last
agent on the path.

All the agents on the path can join or remain in $A$. The newly
assigned good $g$ now belongs to a bundle held by an agent in $A$,
so every agent tracking $g$ can cease doing so, even if they did
not participate in the path. In particular, the starting agent
joins $A$, while no agent leaves $A$. Thus this operation makes
irreversible progress even if it decreases some utilities or
increases the number of unassigned goods.

The other two operations do not dismantle any bundle. A direct
insertion reduces the size of the pool by one, while a cycle
rotation preserves the collection of bundles and gives every
participant a strictly more valuable bundle. To prove termination,
we can therefore compare states lexicographically by the number
of agents in $A$, the number of assigned goods, and total utility.
For the running-time analysis, we use a more quantitative measure:
we count the number of bundles that each agent strictly prefers
to their own. A cycle rotation decreases this count, and we bound
how much the other operations can increase it. Section~\ref{sec:completion}
proves that the operations preserve the required invariants, and
Section~\ref{sec:runtime} gives the resulting polynomial bound.

\subsection{The numerical condition}

We now derive the constraint that determines the approximation
factor. Consider an agent $i$ outside the chosen source component
whose exceptional good is the good $g$ we wish to assign. Write
$p=\pi_i$ and $z=w_i(g_i)$. The agent values every bundle in the
component at most $d_i$ and values the added good at $z$. Because the
agent still holds their original matched good, their utility is at
least $p$. Hence the sufficient condition for all EFX comparisons with
the enlarged bundle is
\[
    d_i+z\leq rp.
\]
There are two contributions to $rd_i=K_i$, namely the initial trim
bound $1+p/r$ and the completion bound $q+D(p)-z$. For each of these
bounds, the preceding inequality is most restrictive at the limiting
values $p=r$ and $z=1$. At $p=r$ we have $D(p)=2/r$. The second
contribution to $K_i$ therefore requires
\[
    \frac{1}{r-1}+\frac{2}{r}-1\leq r(r^2-1).
\]
After clearing the positive denominators, this inequality is
equivalent to
\[
    (r^2+r-1)(r^3-2r^2+2r-2)\geq0.
\]
This is the origin of the cubic equation in the main theorem. The
first factor is positive for $r>1$, so we choose $r$ to be the root of
$r^3-2r^2+2r-2=0$. Taking the reciprocal gives the approximation factor
in Theorem~\ref{thm:main}. In Section~\ref{sec:parameters} we verify this inequality for
all permissible $p,z$, as well as the remaining inequalities needed
by the construction.

We state the sufficient numerical conditions together in the following
theorem. This allows us to prove the combinatorial part of the argument
without repeatedly substituting the particular algebraic value of $r$.

\begin{theorem}[Sufficient numerical conditions]\label{thm:parametric}
Let $3/2<r<2$, put $q=1/(r-1)$, $b=q/r$, $d_0=(1+b)/r$, and
$s_0=d_0-(r-1)r$. Suppose
\begin{equation}\label{eq:parameters}
 \begin{gathered}
  1\le b\le d_0\le r<q<2,\qquad q\le b+1\le r^2,
       \qquad 2b\le 1+r,\\
  0<s_0\le (r-1)r,\qquad \max\{b+s_0,2s_0\}\le2,
       \qquad d_0\le1+(r-1)r .
 \end{gathered}
\end{equation}
For every $r<p<q$ and $(r-1)p<z\le1$, put
\begin{equation}\label{eq:row-numbers}
 D(p)=\frac{1+p/r}{r},\quad s=D(p)-z,\quad
 K=\max\{1+p/r,q+s\},\quad d=K/r .
\end{equation}
Suppose that for every such pair $(p,z)$,
\begin{equation}\label{eq:watch-parameters}
     0\le s\le s_0,\qquad b\le d\le r,\qquad
     s\le (r-1)d,\qquad d+z\le rp .
\end{equation}
Then every finite additive instance as in Theorem~\ref{thm:main}
has a complete $(1/r)$-EFX allocation.
\end{theorem}

For clarity, we summarise the roles of these conditions. The bound
$2b\leq1+r$ ensures that an agent who receives a second good during
the first pass reaches utility at least $b$. The inequalities
$q\leq b+1\leq r^2$ bound the initial trims and the small bundles
formed during completion. The bound $0<s_0\leq(r-1)r$ guarantees that
an agent who retains an exceptional good has no other good violating
the insertion condition. The inequality
$\max\{b+s_0,2s_0\}\leq2$ lets us rule out triples created in the third
pass as the source of an excessively valuable trim. Finally,
$d_0\leq1+(r-1)r$ ensures that the recipient of a matched good from a
second-pass triple receives enough utility even without a second
unassigned good.

The conditions involving $s$ and $d$ are used during completion.
The bound $d\geq b$ allows an agent receiving utility at least $d$
to join $A$ once their exceptional good is permanently assigned.
The inequality $s\leq(r-1)d$ ensures that the insertion condition then
holds for the other potentially releasable goods. The upper bound
$d\leq r$ gives $K\leq r^2$ and ensures that an agent still holding
their original singleton, of price above $r$, already has utility
above $d$. The last inequality, $d+z\leq rp$, is the outside-observer
comparison derived above.

We assume \eqref{eq:parameters} and \eqref{eq:watch-parameters}, with the definitions in \eqref{eq:row-numbers},
throughout Sections~\ref{sec:initial}--\ref{sec:completion} and in the parametric proof in Section~\ref{sec:parametric-proof}. We will also use the following
consequences. For $r<p<q$,
\[
    2<1+\frac{p}{r}<b+1,
    \qquad b=\frac{q}{r}<\frac{2}{r}<D(p)<d_0,
\]
and $q\leq K=rd\leq r^2$.

\section{The first two allocation passes}\label{sec:initial}

We first construct a partial allocation using the matching and the
rescaled valuations of Lemma~\ref{lem:normalise}. The construction consists of two
passes through the agents. In the first, we consider agents whose
matched goods have price at most $r$. We can give each of these agents
an additional good if needed to establish their insertion conditions.
In the second, we consider agents with more expensive matched goods,
for whom such a pairing need not preserve approximate EFX. We use
exchanges of matched goods to deal with these agents.

At the end of the two passes, we will have an approximately EFX
allocation with at most one exception to the insertion condition for
each agent. In addition to this property, the lemma records two facts
needed for the subsequent exchanges. Each three-good bundle contains
two unmatched goods that by themselves provide its owner with
utility at least $d_0$. Also, agents with utility below $d_0$ satisfy
the insertion condition on the entire set of goods left unassigned
by the first pass, not merely the goods left by the second. These
facts will allow us to replace the matched goods in the three-good
bundles later.

\begin{lemma}[Two passes]\label{lem:two-passes}
Under the parameter conditions of Theorem~\ref{thm:parametric},
suppose the matched goods and rescaled valuations satisfy
\eqref{eq:prices}. There is a partial allocation $B$ that assigns
all matched goods. Let $R_1$ be the pool after the first pass,
let $R_0=P(B)$ be the pool after the second, and put $u_i=w_i(B_i)$.
The allocation has the following properties.
\begin{enumerate}
\item For every $i,j\in N$ and $h\in B_j$,
  $w_i(B_j\setminus\{h\})\le\min\{ru_i,b+1\}$.
\item For every $i\in N$, we have $u_i\ge1$, and $u_i<b$ implies
  $B_i=\{a_i\}$.
\item For every $i\in N$, either every good of $R_0$ satisfies
  the insertion condition~\eqref{eq:addition} for $i$, or $B_i=\{a_i\}$ with
  $r<\pi_i<q$ and there is exactly one good $g_i\in R_0$ with
  $w_i(g_i)>(r-1)u_i$. We call $g_i$ the exceptional good for $i$.
  Its value $z_i=w_i(g_i)$ satisfies $(r-1)r<(r-1)\pi_i<z_i\le1$, and
  \begin{equation}\label{eq:first-correlation}
    w_i(h)<d_0-z_i\qquad(h\in R_0\setminus\{g_i\}).
  \end{equation}
  Different agents may have the same exceptional good.
\item Each bundle with three goods consists of a matched good of
  price below $b$ and two original unmatched goods $e,f\in R_1$.
  Its owner $j$ has $w_j(e)+w_j(f)\ge d_0$ and values every good
  of $R_0$ at most $\min\{w_j(e),w_j(f)\}$. Every trim of every other
  bundle has value at most $q$ to every observer. Each agent
  whose utility is below $d_0$ satisfies the insertion condition on all of $R_1$.
\end{enumerate}
\end{lemma}

\begin{proof}
\emph{First pass: agents with matched price at most $r$.}
Initially give each agent $i$ the singleton $\{a_i\}$. Every agent
has utility at least one. For a matched good of price at most $r$,
adding an unmatched good leaves a pair whose two trims are worth at
most $r$ to every agent. We use this observation to establish the
insertion condition for these agents.

Consider each agent $i$ with $\pi_i\leq r$ once, in any order. If there
are no unassigned goods, do nothing. Otherwise let $g$ be a favourite
unassigned good of $i$, breaking ties arbitrarily. Give $g$ to $i$ if
it violates the insertion condition at utility $u_i=\pi_i$. If the
condition holds for $g$, it also holds for all the other unassigned
goods, and we leave $i$'s singleton unchanged.

Suppose $i$ receives $g$, and write $x=w_i(g)$. The failure of the
insertion condition gives
\[
    x>r\pi_i-\max\{b,\pi_i\}
      =\min\{(r-1)\pi_i,r\pi_i-b\}
      \geq r-b,
\]
where the last inequality uses $\pi_i\geq1$ and $b\geq1$. It is
important that an agent who receives a second good no longer has
utility below $b$: we will need all agents below $b$ to hold their
original singletons. This follows from the inequality $2b\leq1+r$,
since
\[
    u_i=\pi_i+x>1+r-b\geq b.
\]
Moreover, because $g$ was a favourite, every good still unassigned
is worth at most $x$ to $i$. The matching bounds give
$x\leq1\leq\pi_i$, and therefore
\[
    x\leq\frac{\pi_i+x}{2}\leq(r-1)(\pi_i+x)=(r-1)u_i.
\]
Together with $u_i\geq b$, this proves the insertion condition for
every remaining good.

Let $R_1$ denote the pool after the first pass. Since the pool only
shrinks, every agent considered in the first pass now satisfies the
insertion condition on all goods of $R_1$. This condition continues
to hold for those goods, even after they are assigned, as long as the
agent's utility does not decrease.

\emph{Second pass: agents with matched price above $r$.}
The preceding pairing rule is no longer necessarily safe for these
agents. For example, at the factor of Theorem~\ref{thm:main}, we have
$r<5/3<q$. An agent whose matched good is worth $5/3$ may value an
unassigned good at one, which exceeds $(r-1)5/3$. However, adding
that good to their singleton can violate the EFX inequality of an
observer with utility one who values the matched good at $5/3$.
Removing the added good from the pair leaves value $5/3>r$ to that
observer.

We overcome this difficulty by exchanging matched goods. An observer
with utility at least $b$ can afford any singleton of price below
$q=rb$ as a trim. Thus we only need to consider observers below $b$,
who still hold their original matched singletons. If such an observer
values a more expensive matched good at least $b$, we can give them
that good alone and use their original matched good in a replacement
bundle. We require two unassigned goods to compensate the original
owner, and we require that their combined value is at least $d_0$
without counting the matched good. This last requirement will be
used again in the third pass.

Consider each agent $j$ with $\pi_j>r$ once, in any order. Such an agent
still holds $\{a_j\}$ when considered, since the exchanges described
below change another agent's bundle only when that agent holds their
original singleton of price below $b\leq r$. If there are fewer than
two unassigned goods, skip $j$. Otherwise choose a favourite
unassigned good $g$ and a favourite good $h$ among the other
unassigned goods, both according to $w_j$. We act only when
\[
    w_j(g)>(r-1)\pi_j
    \qquad\text{and}\qquad
    w_j(g)+w_j(h)\geq d_0.
\]
The first inequality implies $\pi_j<q$, since $w_j(g)\leq1=(r-1)q$.
If there is an agent $i$ who still holds their original first-pass
singleton $\{a_i\}$, with $\pi_i<b$ and $w_i(a_j)\geq b$, choose one
such agent and set
\[
    B_i'=\{a_j\},\qquad B_j'=\{a_i,g,h\}.
\]
Agent $i$ gains utility and reaches at least $b$, so they cannot be
selected as such an observer again. If there is no such agent, set
$B_j'=\{a_j,g\}$ instead. In both cases leave all other bundles
unchanged. Each exchange reassigns matched goods and assigns goods
from the pool; it neither duplicates a good nor returns a matched
good to the pool. Let $R_0$ denote the unassigned goods at the end
of this pass. In particular, $R_0\subseteq R_1$.

\emph{Utilities and the singleton property.}
Every first-pass agent retained or increased their utility. An agent
who received a triple during the second pass has utility at least
$w_j(g)+w_j(h)\geq d_0$. Every other second-pass agent has utility at
least $\pi_j>r\geq d_0$. Thus all utilities are at least one, and every
agent whose utility is below $d_0$ was considered in the first pass.
By the first-pass argument, all goods of $R_1$ still satisfy the
insertion condition for such an agent at their current utility.
Moreover, an agent who ends with utility below $b$ was neither given
a second good nor selected for an exchange. They have retained their
original matched singleton throughout.

\emph{Approximate EFX and the uniform trim bound.}
We check the final allocation according to the possible shapes of
its bundles. The trim of a singleton is empty. Each trim of a
first-pass pair is worth at most $r$ to every agent, which suffices
because every agent's utility is at least one.

Consider a second-pass pair $\{a_j,g\}$. The unmatched good is worth
at most one, and the matched good is worth at most $\pi_j<q$ to every
observer. Thus the EFX inequalities hold for observers with utility
at least $b$, since $q=rb$. An observer $i$ whose final utility is below
$b$ held their original singleton throughout the second pass. If
$w_i(a_j)\geq b$, the rule would have exchanged matched goods instead
of forming this pair. Hence $w_i(a_j)<b\leq r\leq ru_i$, and the
inequalities hold for these observers as well.

Finally, consider a triple. Its matched good has price below $b$,
and both unmatched goods belong to $R_1$. By Lemma~\ref{lem:cheap}, every
trim has value at most $b+1=rd_0$ to every agent. An observer whose
utility is at least $d_0$ therefore satisfies all the EFX inequalities
for this triple. Any observer below $d_0$ satisfies the insertion
condition on both unmatched goods, as established above. The second
part of Lemma~\ref{lem:cheap} gives the required bound $ru_i$ for such an
observer.

We have proved $(1/r)$-EFX. The uniform upper bounds for trims of
singletons, first-pass pairs, second-pass pairs and triples are,
respectively, $0,r,q,b+1$. Since $r<q\leq b+1$, they also prove the
claimed uniform bound $b+1$ on all trim values.

\emph{The remaining goods.}
Every first-pass agent continues to satisfy the insertion condition
on the final pool because their utility did not decrease and
$R_0\subseteq R_1$. An agent $j$ who received a second-pass pair
$\{a_j,g\}$ values every remaining good at most $w_j(g)$, since $g$
was a favourite when selected. The matching bounds give
$w_j(g)\leq\pi_j$, so
\[
    w_j(g)\leq\frac{\pi_j+w_j(g)}{2}\leq(r-1)u_j.
\]
For an agent $j$ who received a triple, every remaining good is worth
at most $w_j(h)$, where $h$ was their second favourite. Therefore
\[
    w_j(h)\leq\frac{w_j(g)+w_j(h)}{2}\leq(r-1)u_j.
\]
Both types of recipient have utility at least $b$. These inequalities
establish the insertion condition for every good of $R_0$.

It remains to consider a second-pass agent $j$ whose singleton was
left unchanged. If the pool was empty when $j$ was considered, or if
their favourite unassigned good satisfied the insertion condition,
there is no violation in $R_0$. Otherwise, writing $z$ for the value
of their favourite good, we have
\[
    (r-1)r<(r-1)\pi_j<z\leq1,
    \qquad r<\pi_j<q.
\]
The second test must have failed, or there was no second good.
Consequently, every other good $h$ unassigned at that visit satisfies
$w_j(h)<d_0-z$. In particular,
\[
    w_j(h)<d_0-z<s_0\leq(r-1)r<(r-1)\pi_j.
\]
As $u_j=\pi_j>r\geq b$, all these other goods satisfy the insertion
condition. If the favourite was subsequently assigned to someone
else, the condition now holds for all of $R_0$. If it remains
unassigned, it is the unique exceptional good $g_j$, with value
$z_j=z$. This proves \eqref{eq:first-correlation} and the asserted uniqueness.

To obtain the final assertion of the lemma, recall that the two
unmatched goods in every second-pass triple were the two most
valuable unassigned goods to their owner at the time of the exchange.
Their combined value is at least $d_0$, and each of them is worth at
least as much to their owner as any good in $R_0$. These are the two
properties that allow us to replace the matched good in such a triple
in the next stage while preserving the utility guarantee and the
insertion conditions of its owner.
\end{proof}

\section{Exchanges in decreasing order of matched price}\label{sec:price-order}

The first two passes give a partial allocation whose trims are all
worth at most $b+1=rd_0$ to every agent. Some agents, however, still
have an exceptional good. For each such agent $i$, we will refine the
uniform threshold $d_0$ using the original matched price $\pi_i$.
If we can assign their exceptional good, we do so; otherwise, we
obtain a stronger upper bound on the value of every other unassigned
good to $i$, together with a stronger upper bound on the value of
every current trim to $i$.

The key to obtaining both bounds is that the matched good in a
second-pass triple is not needed for its owner to retain utility
$d_0$. If a trim of such a triple is an obstruction to giving $i$ a
bundle of smaller value, its matched good may itself be valuable
enough to compensate $i$. We use this observation to perform a third
pass through the agents who still have exceptional goods. The order
in which we perform the exchanges will ensure that the stronger
bounds, once established, are preserved.

\begin{lemma}\label{lem:initial}
Suppose the matching satisfies~\eqref{eq:prices} and the parameters
satisfy Theorem~\ref{thm:parametric}. There is a partial allocation
$B$ assigning all matched goods. It satisfies $(1/r)$-EFX, every
trim is worth at most $b+1$ to every observer, every utility $u_i$
is at least one, and $u_i<b$ implies $B_i=\{a_i\}$.
Write $R=P(B)$. Each agent $i$ satisfies the insertion condition~\eqref{eq:addition}
on all of $R$, or holds $B_i=\{a_i\}$ with $r<\pi_i<q$ and has exactly one
good $g_i\in R$ for which the condition fails. In the latter
case, put $z_i=w_i(g_i)$ and $D_i=(1+\pi_i/r)/r$. Then
$(r-1)\pi_i<z_i\le1$, and
\begin{align}
 w_i(h)&<D_i-z_i &&(h\in R\setminus\{g_i\}),\label{eq:initial-correlation}\\
 w_i(B_j\setminus\{h\})&\le1+\pi_i/r
       &&(j\in N,\ h\in B_j).\label{eq:actual-deletion}
\end{align}
\end{lemma}

\begin{proof}
Let $B$ initially be the partial allocation of Lemma~\ref{lem:two-passes}. As in
that lemma, let $R_1$ and $R_0$ be the pools after the first and second
passes, respectively. For each agent with an exceptional good in
$R_0$, fix that good $g_i$ and its value $z_i=w_i(g_i)$. The bound
\eqref{eq:first-correlation} implies that every other good of $R_0$ is worth less than
\[
    d_0-z_i<s_0\leq(r-1)r<z_i
\]
to $i$. In particular, $g_i$ is their favourite unassigned good for
as long as it remains unassigned, since the pool only shrinks during
this pass.

Throughout the pass we maintain $(1/r)$-EFX, the uniform trim bound
$b+1$, utility at least one for every agent, and the property that
any agent of utility below $b$ holds their original matched singleton.
These properties hold initially by Lemma~\ref{lem:two-passes}. We also maintain
the following distinction between triples. Every triple formed in
the second pass retains its owner and its two original unmatched
goods, whose combined value to that owner is at least $d_0$. Its
matched good may change. A triple formed during the present pass
will keep all three of its goods for the remainder of the pass. Every
triple contains a matched good of price below $b$, and each good in
a two-good bundle is worth at most $q$ to every agent.

\emph{When to perform an exchange.}
Consider the agents with exceptional goods in $R_0$ in nonincreasing
order of their original matched prices $\pi_i$, breaking ties
arbitrarily. In this order, if a later visitor $i$ replaces the
matched good of a second-pass triple, the replacement has price
below $\pi_i/r\leq\pi_v/r$ for each earlier visitor $v$. This will
preserve the earlier bound $rD_v=1+\pi_v/r$ on the trims of that
triple. If
$g_i$ is already assigned when we consider $i$, skip this agent.
Otherwise $i$ still holds $\{a_i\}$: as the operations below show,
an earlier exchange can modify another agent's bundle only when
that agent holds a low-utility original singleton or a second-pass
triple. Neither type can be the bundle of an unvisited agent with
an exceptional good.

We seek to assign $g_i$ while allowing $i$ to receive utility as low
as
\[
    D_i=\frac{1+\pi_i/r}{r}.
\]
To motivate this threshold, suppose that a pair $\{a_i,g_i\}$ would
violate an EFX inequality for agent $k$. The only possible
problematic trim is $\{a_i\}$, since $w_k(g_i)\leq1$ and $u_k\geq1$.
We must then have
\[
    ru_k<w_k(a_i)\leq\pi_i.
\]
In particular, $u_k<\pi_i/r<b$. By our invariant, $k$ holds their
original matched singleton, so $\pi_k=u_k<\pi_i/r$. If we use $a_k$
in a new triple, every trim of that triple is worth at most
$1+\pi_i/r$ to any agent. This is the reason for the choice of
$D_i$: an agent with utility $D_i$ can afford every such trim.

We perform an operation for $i$ if at least one of the following
two conditions holds:
\begin{enumerate}
    \item Some trim of a current bundle has value strictly greater
    than $rD_i$ to $i$.
    \item There is another unassigned good $h\neq g_i$ such that
    $z_i+w_i(h)\geq D_i$.
\end{enumerate}
If neither condition holds, leave $i$'s singleton and the pool
unchanged. Notice that the failure of the first condition gives the
trim bound required in \eqref{eq:actual-deletion}, and the failure of the second gives
the bound on the other unassigned goods required in \eqref{eq:initial-correlation}. We will
show that later operations preserve these bounds whenever $g_i$
remains unassigned.

\emph{The direct pairing operation.}
Whenever at least one of the two conditions holds, first check
whether the pair $\{a_i,g_i\}$ is safe for every agent. It suffices
to check
\[
    w_k(a_i)\leq ru_k\qquad\text{for every agent }k.
\]
Indeed, the other trim is $\{g_i\}$, worth at most one to every agent.
If this check succeeds, simply give $i$ the pair $\{a_i,g_i\}$ and
leave the other bundles unchanged. Agent $i$ gains utility and
retains utility above $r\geq d_0$.

Otherwise, choose an agent $k$ for whom $w_k(a_i)>ru_k$. As above,
$B_k=\{a_k\}$ and $\pi_k=u_k<\pi_i/r<b$. In each of the exchanges
below, agent $k$ receives $\{a_i\}$, reaching utility strictly above
$ru_k\geq r\geq d_0$. We distinguish two cases according to the
values of the current trims to $i$.

\emph{Case 1: a current trim is worth more than $rD_i$ to $i$.}
We first identify the bundle containing this trim. Since
$rD_i=1+\pi_i/r>2$, it cannot be a singleton, whose trim is empty,
or a pair, whose trims are worth at most $q<2$. Consider a triple
formed earlier in the present pass. Its two unmatched goods lie in
$R_0\setminus\{g_i\}$, because $g_i$ is still unassigned. Each is
worth less than $s_0$ to $i$. Since its matched good has price below
$b$, every trim of that triple is worth at most
$\max\{b+s_0,2s_0\}\leq2$ to $i$. Thus it cannot be the triple in
question either.

The only remaining possibility is a triple formed in the second
pass. Write this triple as $\{\ell,e,f\}$, where $\ell$ is its
current matched good and $e,f$ are its two original unmatched goods.
The trim obtained by removing $\ell$ is worth at most two, since
both $e$ and $f$ are unmatched. Consequently the trim whose value
exceeds $rD_i$ contains $\ell$ and one of $e,f$. Since the unmatched
good in that trim is worth at most one, we obtain
\[
    w_i(\ell)>rD_i-1=\frac{\pi_i}{r}>1.
\]
This identifies a matched good that is sufficiently valuable to
$i$. Let $j$ be the owner of the triple. The agents $i,k,j$ are
distinct because $i$ and $k$ hold different singletons while $j$
holds a triple.

We now exchange the three matched goods as follows. Agent $k$
receives $a_i$ alone, agent $i$ receives $\ell$ together with $g_i$,
and agent $j$ receives $a_k$ in place of $\ell$ while retaining both
$e$ and $f$.
\begin{center}
\begin{tabular}{lll}
\hline
Agent & Old bundle & New bundle \\
\hline
$i$ (visitor) & $\{a_i\}$ & $\{\ell,g_i\}$ \\
$k$ (observer) & $\{a_k\}$ & $\{a_i\}$ \\
$j$ (triple owner) & $\{\ell,e,f\}$ & $\{a_k,e,f\}$ \\
\hline
\end{tabular}
\end{center}
All other bundles remain unchanged. The exchange assigns $g_i$
and preserves a unique owner for every previously assigned good.
Agent $i$ receives utility
\[
    w_i(\ell)+z_i>1+(r-1)r\geq d_0.
\]
Agent $j$ may lose utility because the value of $a_k$ need not be
as large as that of $\ell$ to $j$. However, $j$ retains both $e$ and
$f$, whose combined value is at least $d_0$ by Lemma~\ref{lem:two-passes}. Thus all
three participants have utility at least $d_0$ after the exchange.
Notice also that the new matched good in $j$'s triple has price less
than $\pi_i/r$. This last fact is essential for preserving the trim
bounds of agents considered earlier in the pass.

\emph{Case 2: every current trim is worth at most $rD_i$ to $i$.}
Since we are performing an operation, the second condition must
hold. Choose $h$ to be a favourite among the unassigned goods other
than $g_i$. Set
\[
    B_k'=\{a_i\},\qquad B_i'=\{a_k,g_i,h\},
\]
leaving all other bundles unchanged. The only newly assigned goods
are $g_i,h$. The two unmatched goods of the new triple give $i$
utility at least $D_i>b$. Hence all of $i$'s comparisons with the
surviving old bundles hold, because their trims were worth at most
$rD_i$ before the exchange. The comparisons with $i$'s own new bundle
hold by non-negativity. We do not require this triple to give $i$
utility at least $d_0$ from its unmatched goods alone. Accordingly,
we keep it unchanged for the rest of this pass, rather than using
its matched good in a later exchange.

\emph{Preservation of approximate EFX.}
The operations above preserve the asserted bundle shapes. In
particular, the matched good in each triple has price below $b$,
and every good in a pair is worth at most $q$. It follows that the
uniform upper bound $b+1=rd_0$ on trim values continues to hold.
Moreover, each agent whose bundle is changed attains utility at
least $b$. Every agent continues to have utility at least one, and
an agent below $b$ still holds their original matched singleton.

Every agent with utility at least $d_0$ satisfies all EFX
inequalities by the uniform trim bound. Thus it remains to check
the comparisons for two kinds of agents who may have utility below
$d_0$.

First consider an agent whose utility was below $d_0$ at the end of
the second pass. Until selected as an observer $k$ in an exchange,
such an agent keeps their bundle, and every good of $R_1$ satisfies
their insertion condition by Lemma~\ref{lem:two-passes}. Both unmatched goods of
every modified second-pass triple belong to $R_1$, and the same is
true of a new third-pass triple, since $R_0\subseteq R_1$. The second
part of Lemma~\ref{lem:cheap} therefore proves the EFX comparisons with
all such triples for this agent. A new pair using a matched good
from a second-pass triple has both goods worth at most $r$ to every
agent, because that matched good has price below $b\leq r$. A new
direct pair was explicitly checked against every observer when
formed. These observations establish the EFX inequalities for this
first class of agents. Once such an agent is chosen as $k$, their
utility becomes greater than $r\geq d_0$ and the uniform bound
suffices thereafter.

Next consider an agent $v$ who received a new triple in Case~2
and whose utility is below $d_0$. At the time of that exchange we
checked that every old trim was worth at most $rD_v$ to $v$. We
must show that every bundle modified or formed later in the pass
also satisfies the required EFX inequality for $v$ at their utility
of at least $D_v$. This is where the nonincreasing order of original
matched prices is used.

At every later visit to an agent $i$, we have $\pi_i\leq\pi_v$.
If the operation modifies a second-pass triple, the matched good
inserted into that triple has price below $\pi_i/r\leq\pi_v/r$.
Its two unmatched goods still have value at most one each to $v$.
Consequently, every trim of the modified triple is worth at most
\[
    \max\left\{2,1+\frac{\pi_v}{r}\right\}
       =1+\frac{\pi_v}{r}=rD_v
\]
to $v$. Every later third-pass triple contains two unmatched goods
of $R_0$ different from $g_v$, which has already been assigned.
By \eqref{eq:first-correlation}, both are worth less than $s_0$ to $v$. All its trims
are therefore worth at most $\max\{b+s_0,2s_0\}\leq2<rD_v$ to $v$.
Every newly formed pair has trim values at most $q<2<rD_v$. Finally,
$v$ keeps their own triple, so their own utility cannot decrease.
This proves that all their EFX inequalities continue to hold. It
also explains why the nonincreasing order is needed even though
a second-pass triple may be modified more than once.

The two classes exhaust the agents below $d_0$, since all visiting
agents originally had utility above $r\geq d_0$, and every changed
agent except possibly a recipient in Case~2 reaches utility at
least $d_0$. This completes the induction establishing approximate
EFX and the bundle-shape and utility invariants.

\emph{Properties of the final pool.}
Let $R=P(B)$ be the pool at the end of the third pass. We first
show that every agent whose exceptional good has been assigned,
or who had no exceptional good after the second pass, satisfies
the insertion condition on $R$.

The owner of a second-pass triple retains both original unmatched
goods, each worth at least as much to them as any good of $R_0$.
The same property with respect to the final pool holds for the
owner of a third-pass triple: its two unmatched goods were chosen
as their favourite and second favourite. In either type of triple,
every remaining good is worth at most half the owner's current
utility. An agent who received a pair in this pass values the
matched good in that pair at least as much as their exceptional
good: either the matched good is $a_i$, of value $\pi_i>r$, or it
is $\ell$, of value greater than one. The exceptional good is worth
at most one and was their favourite in the pool. Thus every
remaining good is again worth at most half their utility. All these
agents have utility at least $b$. Since $r-1\geq1/2$, their insertion
conditions hold on $R$.

An agent selected as an observer $k$ in an exchange increased
their utility, preserving the insertion condition previously
established on $R_0$. For an unchanged agent who satisfied the
condition after the second pass, the condition continues to hold
because $R\subseteq R_0$. For an unchanged agent whose exceptional
good was assigned to someone else during the present pass, all the
remaining goods satisfy the condition by \eqref{eq:first-correlation}. Thus only agents
whose exceptional goods remain unassigned can have a violation.

Consider such an agent $i$. They still hold $\{a_i\}$, and $g_i$ still
violates the insertion condition, while all other goods of $R$
satisfy it. Neither of the two conditions for performing an
operation held when we considered $i$. Therefore, at that time,
every other unassigned good was worth less than $D_i-z_i$ to $i$,
and every trim was worth at most $rD_i=1+\pi_i/r$ to $i$. The first
bound immediately gives \eqref{eq:initial-correlation}, since the pool only shrinks.

For the second bound, we use the same decreasing-price argument
as for the earlier recipient $v$ above. Every later modified
second-pass triple has a matched good of price below $\pi_i/r$;
every later third-pass triple uses two goods in
$R_0\setminus\{g_i\}$, both worth less than $s_0$ to $i$; and every
later pair has trims worth at most $q<2$. Thus all later changes
preserve the bound $1+\pi_i/r$ on trim values. This establishes
\eqref{eq:actual-deletion} and completes the proof.
\end{proof}

The third pass gives us more than a bound on the number of exceptions.
For an agent $i$ whose exceptional good remains unassigned, we now
know how valuable all the other unassigned goods can be to $i$, and
how much utility $i$ would need in order to satisfy the EFX
inequalities for all the current bundles. The next section uses
these two bounds to complete the allocation.

\section{Completing the allocation}\label{sec:completion}

\subsection{Admissible states}\label{sec:admissible}

In the construction so far, every operation has either reassigned a
matched good or assigned a previously unassigned good. No good has
returned to the pool. During completion, we will sometimes need to
replace a bundle by one containing only some of its old goods and
a newly assigned good. We must therefore keep track of which goods
can become unassigned again.

Let $R$ denote the pool at the end of the construction in
Lemma~\ref{lem:initial}. This set remains fixed throughout the completion
argument. The notation $P(B)$, on the other hand, denotes the current
pool of a partial allocation $B$. We allow goods of $R$ to be
assigned and subsequently returned to the pool. Goods of $M\setminus
R$ must remain assigned, although their owners may change. In
particular, $P(B)\subseteq R$ at every stage.

For each agent $i$ with an exceptional good at the end of
Lemma~\ref{lem:initial}, fix its identity $g_i$ and value $z_i=w_i(g_i)$. Also
write $s_i=D(\pi_i)-z_i$. The lemma gives
\[
    r<\pi_i<q,\qquad (r-1)r<(r-1)\pi_i<z_i\leq1,
\]
and $w_i(h)<s_i$ for every $h\in R\setminus\{g_i\}$. These are
properties of the fixed valuations and the fixed set $R$; they
remain true whether the goods of $R$ are assigned or unassigned.

At the end of the initial construction, every trim is worth at
most $1+\pi_i/r$ to $i$. During completion, however, bundles may
grow. When we insert a good different from $g_i$ into a bundle in
a source component not containing $i$, the value of that old bundle
to $i$ may be as large as $q$, and the added good is worth less than
$s_i$. We therefore use the following bound, as motivated in the
proof overview:
\[
    K_i=\max\left\{1+\frac{\pi_i}{r},q+s_i\right\},
    \qquad d_i=\frac{K_i}{r}.
\]
These quantities are fixed for the remainder of the proof. If all
trims under consideration have value at most $K_i$ to $i$, then a
utility of $d_i$ is sufficient for all the corresponding EFX
inequalities.

We keep a set $W$ of agents whose exceptional goods still need to
be taken into account. Every agent in $W$ has an exceptional good
fixed as above. Membership in $W$ is part of the state of the
construction, rather than something we determine afresh from the
current pool. Indeed, an agent may remain in $W$ even when their
exceptional good is assigned, because that good may subsequently
return to the pool. For brevity, we call agents in $W$ \emph{watchers}
and say that $i\in W$ tracks $g_i$.

For an agent $i\in W$, we maintain the following additional bound:
every trim of every bundle that does not contain $g_i$ has value at
most $K_i$ to $i$. We do not impose this bound on a bundle containing
$g_i$; the ordinary approximate EFX inequalities still apply to that
bundle. When $g_i$ is unassigned, the additional bound applies to
every allocated bundle. In this situation, agent $i$ can receive
utility as low as $d_i$ without violating any comparison with a
surviving old bundle. Their comparisons with newly formed bundles
will be checked separately.

There are two ways for an agent to cease tracking their exceptional
good. First, once their utility reaches $q$, the insertion condition
holds for every good of $R$, because $q\geq b$ and $(r-1)q=1$.
Second, if their exceptional good can never again become unassigned,
they need only satisfy the insertion condition for the other goods
that can return to the pool. A utility of at least $d_i$ suffices
for these goods, since $d_i\geq b$ and $s_i\leq(r-1)d_i$.

To make this second possibility precise, for a state $(B,W)$ let
$u_i=w_i(B_i)$ and define
\[
    A=\{i\in N\setminus W:u_i\geq b\}.
\]
We will construct operations with two properties: no agent already
in $A$ leaves $A$, and no good in a bundle held by an agent of $A$
subsequently becomes unassigned or moves to a bundle outside $A$.
This does not mean that the owners of those goods never change.
We may transfer whole bundles among agents in $A$, including agents
who join $A$ during the operation, and we may enlarge such bundles.
The important property is that none of these goods can return to
the pool. Thus assigning $g_i$ to a bundle of $A$ is sufficient to
ensure that $g_i$ remains assigned permanently.

The goods that can still become unassigned are contained in
\[
    R\setminus\bigcup_{j\in A}B_j.
\]
This set includes the current pool, but it also includes goods that
are currently held by agents outside $A$. For each agent outside
$W$, we maintain the insertion condition on this entire set, not
only on $P(B)$. This ensures that the condition remains valid for
any of these goods that is returned to the pool in a later move.

Finally, we require every agent outside $A$ to hold their original
matched good, possibly together with some goods of $R$. This gives
us a convenient way of replacing one of these bundles: we can
retain its matched good and return some of its other goods to the
pool without returning any good of $M\setminus R$.

We call a state $(B,W)$ \emph{admissible} if it satisfies the following
conditions.
\begin{enumerate}
\item Every good in $M\setminus R$ is assigned, and $B$ is $(1/r)$-EFX for $w$.
\item For every $i\notin A$, there is a set $E_i\subseteq R$
  such that $B_i=\{a_i\}\cup E_i$. For every $i\in W$, we have
  $u_i<q$, and
  \begin{equation}\label{eq:conditional-trims}
    w_i(B_j\setminus\{h\})\le K_i
    \qquad\text{for every $j$ with $g_i\notin B_j$ and every $h\in B_j$}.
  \end{equation}
\item For every $i\notin W$ and every
  $h\in R\setminus\bigcup_{j\in A}B_j$, inequality
  \eqref{eq:addition} holds for $i$ and $h$.
\end{enumerate}
In an admissible state, every agent has utility at least one. This
follows from $u_i\geq b\geq1$ for agents in $A$, and from the presence
of their original matched good for agents outside $A$. In particular,
an agent $i\in W$ is outside $A$ and has utility at least
$\pi_i>r$.

The partial allocation of Lemma~\ref{lem:initial} gives an admissible initial
state. Take $W$ to be the set of agents with an exceptional good at
the end of that lemma. Every agent in $W$ then holds their original
matched singleton, with utility $\pi_i<q$. Every other agent outside
$A$ has utility below $b$ and also holds their original matched
singleton. The initial trim bound \eqref{eq:actual-deletion} is at most $K_i$ for
every $i\in W$, so the conditional bound \eqref{eq:conditional-trims} holds. Finally,
every agent outside $W$ satisfies the insertion condition on all of
$R$ by Lemma~\ref{lem:initial}, and therefore on the subset of potentially
releasable goods. The remaining requirements are precisely the
approximate EFX and coverage properties of Lemma~\ref{lem:initial}.

\subsection{Progress and the graph for an unassigned good}

A reassignment that takes apart a bundle may return several goods
to the pool. We will ensure that every such reassignment brings a
new agent into $A$, while every agent already in $A$ remains
there. The other operations either assign one additional good
without dismantling a bundle or rotate the existing bundles so
that every participant strictly increases their utility. We use
these three kinds of progress to prove that the completion
procedure terminates.

\begin{lemma}[Completion]\label{lem:complete}
Starting from the partial allocation of
Lemma~\ref{lem:initial}, the completion procedure
below maintains admissibility and terminates with a complete
$(1/r)$-EFX allocation.
\end{lemma}

\begin{proof}
Consider an admissible state $(B,W)$. If $P(B)=\varnothing$,
we already have a complete allocation. Otherwise, choose any good
$g\in P(B)$. We describe an operation that produces another
admissible state and then show that repeated application of these
operations terminates.

\emph{Receiving thresholds.}
For the chosen good $g$, let
\[
    W_g=\{i\in W:g_i=g\}.
\]
These are precisely the agents still tracking the good that we are
trying to assign. We call them \emph{active} for this operation.
The set $W_g$ can contain several agents, since their exceptional
goods need not be distinct.

We want an agent who receives another bundle to be able to join or
remain in $A$. There are three cases. An agent $i\notin W$ can
preserve their insertion conditions by retaining at least their
current utility, and can join $A$ if their utility is at least $b$.
Thus a utility of $\max\{b,u_i\}$ suffices for these purposes.

For a nonactive watcher $i\in W\setminus W_g$, we cannot rely on
assigning $g_i$ permanently in the present operation. Instead, we
require the new bundle to give them utility at least $q$. At that
utility, the insertion condition holds for every good of $R$,
including $g_i$, and the agent can leave $W$.

Finally, suppose $i\in W_g$. The good $g_i=g$ is currently unassigned,
so the conditional trim bound \eqref{eq:conditional-trims} applies to all old bundles.
It is sufficient for $i$ to receive utility $d_i=K_i/r$ to satisfy
all comparisons with surviving old bundles. If the operation puts
$g$ in a bundle of $A$, then every good that remains potentially
releasable is different from $g_i$ and hence has value below $s_i$
to $i$. The inequalities $d_i\geq b$ and $s_i\leq(r-1)d_i$ imply
that all these goods satisfy the insertion condition at utility
$d_i$. Provided we also check the new-bundle comparisons, we can
therefore remove $i$ from $W$ at this lower utility.

These considerations motivate the receiving thresholds
\[
    t_i=\begin{cases}
        \max\{b,u_i\}, & i\notin W,\\
        q, & i\in W\setminus W_g,\\
        d_i, & i\in W_g.
    \end{cases}
\]
Draw an edge $i\to j$ between distinct agents if $w_i(B_j)>t_i$.
The strict inequality ensures that a transfer strictly improves
the recipient's utility whenever their threshold is at least their
current utility. This is the case for every agent outside $W_g$,
because $u_i<q$ for watchers. For an active watcher, however, the
threshold $d_i$ can be smaller than their current utility. We will
use such transfers only in an operation that assigns $g$
permanently.

\emph{Bounds for agents outside a source component.}
Choose a source strongly connected component $C$ of this graph.
For every $i\notin C$ and $j\in C$, the absence of an edge from $i$
to $j$ gives $w_i(B_j)\leq t_i$. Thus

\begin{equation}\label{eq:outside-source}
 w_i(B_j\cup\{g\})\le
 \begin{cases}
   ru_i, & i\notin W,\\
   q+s_i, & i\in W\setminus W_g,\\
   d_i+z_i, & i\in W_g.
 \end{cases}
\end{equation}

For an agent $i\notin W$, the insertion condition holds for $g$,
since $g$ is currently unassigned. Consequently,
$\max\{b,u_i\}+w_i(g)\leq ru_i$. For a nonactive watcher, $g\neq g_i$,
so $w_i(g)<s_i$. In this case
\[
    q+s_i\leq K_i\leq r^2<r\pi_i\leq ru_i.
\]
This bounds both the value required for approximate EFX and the
value required by the conditional invariant. For an active
watcher, we use the last inequality in \eqref{eq:watch-parameters} to obtain
\[
    d_i+z_i\leq r\pi_i\leq ru_i.
\]
Hence every agent outside $C$ can afford every trim of
$B_j\cup\{g\}$ at their current utility, for any $j\in C$. Moreover,
for every outside watcher with $g_i\neq g$, the enlarged bundle
has value at most $K_i$, so their conditional trim bound is also
satisfied whenever it is required. These conclusions remain true
for any subset of $B_j\cup\{g\}$, by non-negativity.

We now distinguish three cases, according to the size of $C$ and
whether it contains an active watcher.

\emph{Case 1: $C$ is a singleton.}
Write $C=\{k\}$ and give $g$ to $k$, leaving all other bundles
unchanged. The outside-observer bounds \eqref{eq:outside-source} establish every EFX
comparison with the enlarged bundle for agents other than $k$.
Agent $k$'s utility cannot decrease, and their comparisons with
their own bundle hold by non-negativity. Hence the new allocation
is $(1/r)$-EFX.

If $k$ is a watcher, remove $k$ from $W$ if their new utility is at
least $q$, and otherwise keep $k$ in $W$. An agent reaching $q$
satisfies the insertion condition on all of $R$. All agents who
were already in $A$ remain in $A$, since no utility decreases.
Their goods remain in bundles of $A$. In particular, the set of
potentially releasable goods cannot grow. The insertion conditions
for agents previously outside $W$ therefore continue to hold.
Every agent still outside $A$ retains their original matched good;
if such an agent received an additional good, that good is in $R$.
No good of $M\setminus R$ has become unassigned.

To check the conditional trim bound \eqref{eq:conditional-trims}, consider an agent
$i$ who remains in $W$. The only new bundle is $B_k\cup\{g\}$. If
$i$ tracks $g$, this bundle contains $g_i$, so the additional
bound is not required for it. If $i$ tracks a different good and
$i\neq k$, its entire value to $i$ is at most $q+s_i\leq K_i$ by
\eqref{eq:outside-source}. If $i=k$ and $g_i\neq g$, the agent's new utility is
below $q\leq K_i$, since they remain in $W$. Thus the bound also
holds for each trim of their enlarged bundle. All other bundles
are unchanged. In particular, the utility of every remaining
watcher is below $q$.

The resulting state is admissible. No agent leaves $A$ and the
number of assigned goods increases by one.

\emph{Case 2: $|C|\geq2$ and $C\cap W_g=\emptyset$.}
Every agent in $C$ has a receiving threshold at least their current
utility. Choose a directed cycle in $C$ and give each participating
agent their successor's old bundle. By the definition of the edges,
every participating agent strictly increases their utility. Every
other agent retains their old utility.

Remove all the participants from $W$. Each participant previously
outside $W$ has new utility greater than $\max\{b,u_i\}$; each
participant previously in $W$ has new utility greater than $q$.
Thus all participants belong to $A$ after the operation. Moreover,
the participants removed from $W$ satisfy the insertion condition
on all of $R$ because their new utilities exceed $q$.

As the collection of bundles has not changed and no utility has
decreased, the allocation is still $(1/r)$-EFX. The conditional
trim bounds for the remaining watchers are preserved because every
bundle remains unchanged as a set. A remaining watcher keeps their
old bundle and utility below $q$. Every agent still outside $A$
keeps their original matched good, and all goods of $M\setminus R$
remain assigned. No agent leaves $A$, and every bundle transferred
is received by an agent in $A$. Consequently, no good moves from
an old bundle of $A$ to a bundle outside $A$. The potentially
releasable set can only shrink. Since the utilities of previously
untracked agents do not decrease, their insertion conditions
continue to hold on that set.

The new state is admissible. The pool and the collection of bundles
are unchanged. Hence the strict improvement of every participant
also strictly improves their rank among the current bundles.

\emph{Case 3: $|C|\geq2$ and $C\cap W_g\neq\emptyset$.}
Here we cannot simply rotate along an arbitrary directed cycle.
An active watcher may receive a bundle of value above $d_i$ but
below their current utility, while $g$ remains unassigned. Instead,
we will place $g$ permanently in a bundle of $A$ and ensure that
at least one new agent enters $A$.

\medskip
\noindent\emph{Choosing the starting bundle.}
We need to choose an agent $j_0\in C\setminus A$ whose bundle we can
replace. The choice depends on whether some member of $C$ has
utility below $b$. If there is such a member, choose them as $j_0$.
Their original matched good has price at most their utility and
hence below $b$. This will let us apply Lemma~\ref{lem:cheap} to small
replacement bundles. If every member of $C$ has utility at least
$b$, choose any $j_0\in C\cap W_g$. In this case the matched good
of $j_0$ may have price above $b$, but its price is below $q=rb$,
which every member of $C$ can afford as a trim.

In either case $j_0\notin A$, so the admissibility conditions give
\[
    B_{j_0}=\{a_{j_0}\}\cup E
    \qquad\text{for some }E\subseteq R.
\]
The matched good $a_{j_0}$ lies outside $R$ and must remain assigned.
We form a replacement bundle from $a_{j_0}$, the unassigned good
$g$, and a subset of $E$.

\medskip
\noindent\emph{Constructing a replacement bundle.}
We want some member of $C$ to receive the replacement, but we also
need to control its trims for every other member. We achieve both
by choosing an inclusion-minimal set $H\subseteq E$ such that
\[
    S=\{a_{j_0},g\}\cup H
    \quad\text{satisfies}\quad w_k(S)>t_k
    \quad\text{for some }k\in C.
\]
Such a set exists. Since $C$ is a nonsingleton strongly connected
component, $j_0$ has an incoming edge from another member of $C$.
That member values $B_{j_0}$ strictly above their threshold and
therefore also values $B_{j_0}\cup\{g\}$ strictly above their threshold.
Thus $H=E$ is a possible choice. Furthermore, if $j_0$ was chosen
from $W_g$ in the second case above, the empty set already works:
\[
    w_{j_0}(\{a_{j_0},g\})\geq\pi_{j_0}>r
        \geq d_{j_0}=t_{j_0}.
\]
In this case inclusion-minimality forces $H=\emptyset$.

Fix a member $k\in C$ witnessing the acceptance of the chosen
replacement $S$. The minimality of $H$ is with respect to the
existence of \emph{any} accepting member of $C$, not merely this
particular recipient $k$. Therefore
\[
    w_i(S\setminus\{h\})\leq t_i
       \qquad(i\in C,\ h\in H).
\]
Otherwise $H\setminus\{h\}$ would also have satisfied the defining
property of $H$. This observation controls every trim obtained by
removing a retained good from $H$. To establish all EFX inequalities
for $S$, we must also bound the trims obtained by removing the
matched good $a_{j_0}$ or the newly assigned good $g$.

\medskip
\noindent\emph{Trims of the replacement.}
For every observer outside $C$, the required inequalities follow
from the previously established outside-observer bounds, because
$S\subseteq B_{j_0}\cup\{g\}$. In particular, a nonactive watcher
outside $C$ values the whole of $S$ at most $q+s_i\leq K_i$. We
therefore concentrate on observers inside $C$, distinguishing three
possible sizes of $H$.

\begin{itemize}
\item Suppose $H=\emptyset$. The replacement is the pair
$\{a_{j_0},g\}$. If $j_0$ was chosen with utility below $b$, then
$\pi_{j_0}<b\leq r$, and both trims of the pair are worth at most
$r$ to every observer. Every observer has utility at least one.
Otherwise $j_0$ is an active watcher with matched price below $q$,
and all members of $C$ have utility at least $b$. In this case both
trims are worth at most $q=rb$ to every observer. Thus every member
$i$ of $C$ satisfies the required bound $ru_i$ for the new pair.
Moreover, every watcher $i\in C$ values its trims at most
$q\leq K_i$.

\item Suppose $H=\{h\}$. The starting agent must have utility below
$b$, because the choice of an active starting agent when all
utilities are at least $b$ forces $H=\emptyset$. Therefore
$\pi_{j_0}<b$ and $S=\{a_{j_0},g,h\}$ contains a low-price matched
good and two unmatched goods.

For $i\in C\setminus W$, both $g$ and $h$ were potentially
releasable before the operation. Indeed, $g$ was unassigned and
$h$ belonged to the bundle of $j_0\notin A$. Both goods therefore
satisfy the insertion condition for $i$. Lemma~\ref{lem:cheap} gives the
required bound $ru_i$ for every trim of $S$.

For an active watcher $i\in C\cap W_g$, the minimality of $H$ gives
$w_i(S\setminus\{h\})\leq d_i$. Since $h\neq g_i=g$ and $h\in R$,
we have $w_i(h)<s_i$. Consequently,
\[
    w_i(S)\leq d_i+w_i(h)<d_i+s_i\leq q+s_i\leq K_i.
\]
Thus every trim of $S$ has value at most $K_i$, even if this agent's
utility subsequently decreases to $d_i$.

Now consider a nonactive watcher $i\in C\cap(W\setminus W_g)$.
If $g_i\notin S$, both unmatched goods in $S$ are worth less than
$s_i$ to $i$. Every trim is therefore worth at most
\[
    \max\{b+s_i,2s_i\}\leq q+s_i\leq K_i,
\]
where the middle inequality uses $b<q$ and
$s_i\leq(r-1)d_i\leq d_i\leq r<q$. If instead $g_i\in S$, the
conditional bound \eqref{eq:conditional-trims} is not required for this bundle. The
ordinary EFX bound follows from Lemma~\ref{lem:cheap}:
\[
    w_i(S\setminus\{x\})\leq b+1\leq r^2
       <r\pi_i\leq ru_i\qquad(x\in S).
\]
A nonactive watcher cannot lose utility in the transfer to be
performed below, so this last bound suffices.

\item Suppose $|H|\geq2$. In this case we bound the value of the
whole replacement by adding one retained good back to a trim
controlled by minimality. For $i\in C\setminus W$, any $h\in H$
was held by $j_0\notin A$ before the move, so it satisfies the
insertion condition for $i$. Therefore
\[
    w_i(S)\leq t_i+w_i(h)
       =\max\{b,u_i\}+w_i(h)\leq ru_i.
\]
For a watcher $i\in C\cap W$, there is a good $h\in H$ different
from $g_i$, since $H$ contains at least two goods. This good is
worth less than $s_i$ to $i$. Using the fact that every watcher's
receiving threshold is at most $q$, we obtain
\[
    w_i(S)\leq t_i+w_i(h)<q+s_i\leq K_i.
\]
These bounds on the whole replacement also bound every possible
trim of it.
\end{itemize}

We have established the EFX comparisons with $S$ at the old
utilities of all observers. In addition, every active watcher in
$C$ values each trim of $S$ at most $K_i=rd_i$, so these comparisons
will hold even if their utility falls to $d_i$. For every watcher
who continues to track an exceptional good not in $S$, the
conditional trim bound is also satisfied for $S$.

\medskip
\noindent\emph{Reassigning bundles along a path.}
Since $j_0,k\in C$, there is a simple directed path from $j_0$ to
$k$ within $C$. Give each agent along the path except the last the
old bundle of their successor, and give the last agent $k$ the
replacement $S$. If $j_0=k$, the path has length zero and only
$B_{j_0}$ is replaced. The other bundles are unchanged. The only
old bundle not transferred intact is $B_{j_0}$; its matched good
is retained in $S$, and precisely the goods $E\setminus H$ return
to the pool. Thus all the resulting bundles are pairwise disjoint
and every good of $M\setminus R$ remains assigned.
Figure~\ref{fig:path} depicts the reassignment.

\begin{figure}[htb]
  \centering
  \begin{tikzpicture}[x=1.35cm,y=.50cm,>=Stealth]
    \node[anchor=east] at (0,0) {Path};
    \node[anchor=east] at (0,-1.15) {Old bundle};
    \node[anchor=east] at (0,-2.3) {New bundle};
    \node (a) at (.8,0) {$j_0$};
    \node (b) at (1.85,0) {$j_1$};
    \node (c) at (2.9,0) {$j_2$};
    \node (d) at (3.95,0) {$k$};
    \draw[->] (a) -- (b);
    \draw[->] (b) -- (c);
    \draw[->] (c) -- (d);
    \node at (.8,-1.15) {$B_{j_0}$};
    \node at (1.85,-1.15) {$B_{j_1}$};
    \node at (2.9,-1.15) {$B_{j_2}$};
    \node at (3.95,-1.15) {$B_k$};
    \node at (.8,-2.3) {$B_{j_1}$};
    \node at (1.85,-2.3) {$B_{j_2}$};
    \node at (2.9,-2.3) {$B_k$};
    \node at (3.95,-2.3) {$S$};
  \end{tikzpicture}
  \caption{Reassignment along a path from $j_0$ to $k$. Each agent
  before $k$ receives the next agent's old bundle. Agent $k$
  receives $S$, which contains $g$ and part of $B_{j_0}$. The goods
  in $B_{j_0}$ omitted from $S$ return to the pool.}
  \Description{Four agents on a directed path, with two rows giving
  the old and new bundles of the agents. Every agent except the last
  receives the old bundle of its successor. The last receives the replacement bundle S.}
  \label{fig:path}
\end{figure}

Let $B'$ be the resulting partial allocation and write $u_i'$ for
the new utility of agent $i$. Define $W'$ by removing from $W$ all
the path participants and every agent of $W_g$, including the
active watchers who did not participate in the path. Let $A'$ be
the set defined from $(B',W')$ in the same way as $A$ was defined
from $(B,W)$.

Every participant receives a bundle worth strictly more than their
receiving threshold, either by following an edge or by receiving
$S$. Thus previously untracked participants have
$u_i'>\max\{b,u_i\}$, participating nonactive watchers have
$u_i'>q$, and participating active watchers have $u_i'>d_i\geq b$.
All of them belong to $A'$. A member of $W_g$ who does not
participate in the path keeps their old bundle and hence their old
utility, which is at least $\pi_i>r\geq d_i\geq b$. Such an agent
also belongs to $A'$. No agent in $A$ leaves it. In particular,
$A\subseteq A'$ and $j_0\in A'\setminus A$.

We now verify admissibility for the new state. This requires three
checks, corresponding to the three admissibility conditions.

\medskip
\noindent\emph{Approximate EFX.}
Only active watchers who participate in the path can receive lower
utility. For each such agent $i$, the good $g_i=g$ was unassigned
before the operation. Thus every old bundle lacked $g_i$, and the
conditional bound \eqref{eq:conditional-trims} implies that every old trim is worth at
most $K_i$ to $i$. Since $u_i'>d_i=K_i/r$, all comparisons with
surviving old bundles hold. Their comparisons with $S$ follow from
the bound $K_i$ established above for active watchers inside $C$.
Every other agent keeps or increases their utility and therefore
retains their EFX comparisons with the surviving old bundles.
Their comparisons with $S$ hold by the bounds above at their old
utilities. The resulting allocation is therefore $(1/r)$-EFX.
Together with the coverage and disjointness already checked, this
establishes the first admissibility condition.

\medskip
\noindent\emph{Bundle structure and the conditional trim bound.}
Every agent still outside $A'$ is outside the path and keeps their
old bundle. Since $A\subseteq A'$, they held their original matched
good before the move and still hold it, accompanied only by goods
of $R$. Every remaining watcher retains their old utility below
$q$, since all participating watchers have been removed from $W$.
For any such watcher $i$, the conditional trim bound continues to
hold for every surviving old bundle: its contents did not change.
The bounds for the replacement $S$ established above also imply
\eqref{eq:conditional-trims} whenever $g_i\notin S$. Thus the second admissibility
condition holds.

\medskip
\noindent\emph{Insertion conditions.}
It remains to justify that removing the indicated agents from $W$
was safe for subsequent operations. The only bundle taken apart
was that of $j_0\notin A$. Every surviving old bundle that changes
owner is received by a member of $A'$, and all agents of $A$
remain in $A'$. Hence every good previously held by an agent of
$A$ remains in a bundle of $A'$. Equivalently,
\[
    R\setminus\bigcup_{j\in A'}B_j'
       \subseteq R\setminus\bigcup_{j\in A}B_j.
\]
Thus the set of potentially releasable goods can only shrink.
Every agent previously outside $W$ keeps or increases their
utility, so their insertion conditions continue to hold on the
new set.

An agent removed from $W\setminus W_g$ has new utility greater
than $q$, and therefore satisfies the insertion condition for all
of $R$. Finally, consider an agent $i\in W_g$. The good $g_i=g$
now belongs to $S=B_k'$, and $k\in A'$. In particular, every good
of $R$ still outside the bundles of $A'$ is different from $g_i$
and has value below $s_i$ to $i$. Since $u_i'\geq d_i\geq b$ and
$s_i\leq(r-1)d_i$, the insertion condition holds for each of these
goods at utility $u_i'$. This argument also applies to the active
watchers who did not participate in the path. It proves that
removing all of $W_g$ from $W$ is justified, and establishes the
third admissibility condition.

\medskip
\noindent\emph{Progress.}
We have constructed an admissible state in which every agent of
$A$ remains in $A'$ and the starting agent $j_0$ joins $A'$. Thus
$|A'|>|A|$, regardless of how many goods return to the pool.

Every iteration strictly increases
$(|A|,m-|P(B)|,\sum_i u_i)$ lexicographically. The initial state
is admissible, and there are only finitely many allocations and
watcher sets. Consequently the loop terminates with an empty
pool. The implementation of these operations is given in
Algorithm~\ref{alg:completion}. In Section~\ref{sec:runtime} we bound
the number of operations using an integer-valued measure of the
agents' rankings of the current bundles.
\end{proof}

\section{Proof of the main theorem}\label{sec:main-proof}

\subsection{The parametric guarantee}\label{sec:parametric-proof}

\begin{proof}[Proof of Theorem~\ref{thm:parametric}]
Apply the positive matching reduction of Lemma~\ref{lem:positive-core}.
If no agents remain in the residual instance, the reduction already gives
a complete EFX allocation. Otherwise, apply Lemma~\ref{lem:normalise} to
the residual instance, then construct the partial allocation of
Lemma~\ref{lem:two-passes} and refine it using
Lemma~\ref{lem:initial}. The completion procedure of
Lemma~\ref{lem:complete} yields a complete $(1/r)$-EFX allocation for
the rescaled valuations.
Dividing the inequalities of each agent $i$ by $\lambda_i$ gives the
same guarantee for their original valuation. Lemma~\ref{lem:positive-core}
then extends the allocation to the full instance, including every
comparison that removes a good worth zero to the observer.
\end{proof}

\subsection{Choice of the approximation factor}\label{sec:parameters}

\begin{proof}[Proof of Theorem~\ref{thm:main}]
By Theorem~\ref{thm:parametric}, it remains to choose $r$ and verify the
sufficient numerical conditions. We begin with the condition that
is most restrictive for our construction. Recall that an agent
outside the chosen source component who tracks the unassigned good
being inserted must satisfy $d+z\leq rp$, where $p$ is their
original matched price and $z$ is the value of their exceptional
good. The parameters must satisfy this inequality for every
\[
    r<p<q,\qquad (r-1)p<z\leq1.
\]
By \eqref{eq:row-numbers}, the utility threshold $d$ depends on $p$ and $z$ through
\[
    D(p)=\frac{1+p/r}{r},\quad
    s=D(p)-z,\quad
    rd=K=\max\{1+p/r,q+s\}.
\]
Consequently,
\[
    r(d+z)=\max\left\{
        1+\frac{p}{r}+rz,
        \ q+D(p)+(r-1)z
    \right\}.
\]
Both expressions on the right increase with $z$, while the required
upper bound $r^2p$ is independent of $z$. Since $z\leq1$, it suffices
to prove the two inequalities at $z=1$, namely
\[
    1+\frac{p}{r}\leq r(rp-1),\qquad
    q+D(p)-1\leq r(rp-1).
\]
In each inequality, the right side grows faster with $p$ than the
left side: the respective differences are increasing affine
functions with slopes $r^2-1/r$ and $r^2-1/r^2$, both positive for
$r>1$. Hence it suffices to check their limiting values at $p=r$,
even though the allowed price interval is open at this endpoint.
At $p=r$ we have $D(r)=2/r$, so the two inequalities reduce to
\[
    2\leq r(r^2-1),\qquad
    q+\frac{2}{r}-1\leq r(r^2-1).
\]

We choose $r$ to make the second inequality tight and then verify
that the first inequality also holds. Substituting $q=1/(r-1)$
and clearing the positive denominators gives the identity
\[
 r(r-1)\bigl[r(r^2-1)-q-2/r+1\bigr]
     =(r^2+r-1)(r^3-2r^2+2r-2).
\]
The factors $r(r-1)$ and $r^2+r-1$ are positive for $r>1$, so the
equality is equivalent to $f(r)=0$ for $f(x)=x^3-2x^2+2x-2$.
Since $f'(x)=3(x-2/3)^2+2/3>0$ and
\[
       f(77/50)=-1367/125000<0,\qquad f(31/20)=151/8000>0 ,
\]
there is exactly one real root and $77/50<r<31/20$. The first
comparison also holds because its margin $r^3-r-2$ increases throughout this
interval and equals $14033/125000>0$ at $77/50$. This proves
$d+z\le rp$ throughout the required domain.

Substituting $r=1/\alpha$ in $f(r)=0$ and multiplying by
$-\alpha^3$ gives $2\alpha^3-2\alpha^2+2\alpha-1=0$. Zero is not a
root, so uniqueness of $r$ gives uniqueness of $\alpha$. Equivalently,
$\alpha=T/(T+1)$ for the real $T>1$ satisfying $T^3=T^2+T+1$, by
substitution.

For the decimal guarantee in the title, observe that
\[
    f\left(\frac{10000}{6477}\right)
    =\frac{190473334}{271720053333}>0.
\]
Since $f$ is strictly increasing and $f(r)=0$, this gives
$r<10000/6477$ and hence $\alpha>0.6477$.

The preceding argument explains the choice of $r$ and establishes
the limiting inequality of the construction. We must still check
that all the earlier construction and completion requirements are
satisfied at this value. For this purpose, the rational interval
$77/50<r<31/20$ obtained above is sufficient. Substituting its
endpoints into $r-1$, $r(r-1)$ and the definitions
\[
    q=\frac{1}{r-1},\qquad b=\frac{1}{r(r-1)},\qquad
    d_0=\frac{1+b}{r}
\]
gives the following rational lower and upper bounds.
\[
\begin{array}{l|ccccc}
 &r-1&r(r-1)&q&b&d_0\\ \hline
 \text{lower}&54/100&83/100&9/5&7/6&139/100\\
 \text{upper}&55/100&86/100&93/50&121/100&36/25 .
\end{array}
\]
For instance, $r(r-1)$ lies between
$(77/50)(54/100)$ and $(31/20)(55/100)$. Taking reciprocals
gives the stated bounds on $b$. Dividing $1+7/6$ by $31/20$ and
$1+121/100$ by $77/50$ bounds $d_0$. Consequently
$1<b<d_0<r<q<2$, $q<b+1<221/100<(77/50)^2$, and
$2b<242/100<1+r$. The table also gives
\[
  0<s_0=d_0-r(r-1)<61/100<r(r-1),\qquad
  b+s_0<182/100,\qquad 2s_0<122/100 .
\]
Finally $1+r(r-1)>183/100>d_0$.
These are all the conditions in~\eqref{eq:parameters}.

We next verify the remaining conditions in \eqref{eq:watch-parameters} for every
permissible pair $(p,z)$. Since $r<p<q$, we have
$2/r<D(p)<d_0$. Together with $z\leq1$ and $r<2$, this gives
$s=D(p)-z>2/r-1>0$. The inequality $z>(r-1)p>(r-1)r$ also gives
$s<d_0-(r-1)r=s_0$.

To establish the upper bound on $d=K/r$, we need a slightly
stronger bound on $s$ than $s_0$. For this we use the relationship
between the original price $p$ and the value of the exceptional
good $z$, namely $z>(r-1)p$.
Since $1/r^2<1/(77/50)^2<54/100<r-1$, the function
\[
              D(p)-(r-1)p=1/r+p\big(1/r^2-(r-1)\big)
\]
decreases with $p$. Hence
$s<2/r-r(r-1)<13/10-83/100=47/100$. It follows that
\[
  rb=q\le K\le\max\{1+b,\ q+47/100\}<233/100<(77/50)^2<r^2 .
\]
Thus $b\le d\le r$. Moreover
$s<s_0<61/100<(54/100)(7/6)<(r-1)b\le(r-1)d$.
Together with the previously established inequality $d+z\leq rp$,
this proves every condition in \eqref{eq:watch-parameters}. All the hypotheses of
Theorem~\ref{thm:parametric} are therefore satisfied. We obtain a complete
$\alpha$-EFX allocation for $\alpha=1/r$, the unique real root of
$2\alpha^3-2\alpha^2+2\alpha-1=0$, as claimed.
\end{proof}

\begin{proof}[Proof of Corollary~\ref{cor:rational}]
Since $r<31/20$, the factor $1/r$ strictly exceeds $20/31$.
Every rival bundle after one deletion has non-negative value.
The same allocation therefore satisfies all the inequalities at $20/31$.
\end{proof}

\section{The allocation algorithm}\label{sec:algorithm}
We now give the pseudocode for the construction proved correct
in the preceding sections. The algorithm first applies the
positive matching reduction and computes the matching and
rescaled valuations on the residual instance. It then constructs
the initial partial allocation, performs the decreasing-price
exchanges, and completes the allocation. The last three steps
are implemented by the subroutines \textsc{InitialPasses},
\textsc{ResolveExceptions}, and \textsc{CompleteAllocation},
respectively.

\subsection{The complete algorithm and the computation of prices}\label{sec:algorithm-overview}

Algorithm~\ref{alg:allocation} takes the matrix of singleton values as
input. Throughout the algorithms, $r=1/\alpha$, $q=1/(r-1)$,
$b=q/r$, and $d_0=(1+b)/r$. We write $P(B)$ for the current pool and
$u_i=w_i(B_i)$ for the current utility. An assignment to several
bundles on one line uses their old contents throughout. A condition
on $P(B)$ in a loop header is tested when the agent is visited.
All ties and unspecified choices may be resolved in input order.

\begin{algorithm}[htbp]
\SetAlgoLined
\DontPrintSemicolon
\KwIn{A nonempty set $N$ of agents, a set $M$ of goods
with $|M|>|N|$, and non-negative additive valuations $v_i(g)$.}
\KwOut{A complete $\alpha$-EFX allocation $X$.}
$F\gets$ a maximum-cardinality matching on the edges $v_i(g)>0$\;
Orient the edges of $F$ from goods to agents and all other positive edges from agents to goods\;
$(L,T)\gets$ the agents and goods reachable from the unmatched agents\;
$I\gets N\setminus L$, \quad $G\gets M\setminus T$\;
$X_i\gets\{h\}$ if $(i,h)\in F$, and $X_i\gets\emptyset$ otherwise, for each $i\in L$\;
\If{$I=\emptyset$}{
  Choose an unmatched $i\in L$ and set $X_i\gets G$\;
  \Return $X$\;
}
$(a,\pi,w)\gets\textsc{PriceMatching}(I,G,v)$\;
$B\gets\textsc{InitialPasses}(I,G,a,\pi,w)$\;
$(B,W,(g_i)_{i\in W})\gets\textsc{ResolveExceptions}(B,a,\pi,w)$\;
$(X_i)_{i\in I}\gets\textsc{CompleteAllocation}(B,W,(g_i)_{i\in W},a,\pi,w)$\;
\Return $X$\;
\caption{\textsc{Allocate} computes a complete $\alpha$-EFX allocation.}
\label{alg:allocation}
\end{algorithm}
\FloatBarrier

The first part of Algorithm~\ref{alg:allocation} implements the positive
matching reduction of Lemma~\ref{lem:positive-core}. If the reduction
does not already give a complete allocation, the residual instance
admits a matching saturating its agents on positive-valued edges.
We next describe how to compute the matching and the scaling
factors required by Lemma~\ref{lem:normalise}.

\paragraph{Computing the prices.}
The subroutine $\textsc{PriceMatching}(I,G,v)$ first obtains a
maximum-product matching on positive edges saturating $I$. Write
$a_i$ for the good matched to $i$ and $U=G\setminus\{a_i:i\in I\}$.
This set is nonempty by Lemma~\ref{lem:positive-core}. One vertex
$*$ suffices to represent the dummy agents of
Lemma~\ref{lem:normalise}. On $V=I\cup\{*\}$ use arc weights
\[
 \rho_{ij}=\frac{v_i(a_j)}{v_i(a_i)}\quad(i,j\in I),\qquad
 \rho_{i*}=\max_{h\in U}\frac{v_i(h)}{v_i(a_i)}\quad(i\in I),\qquad
 \rho_{*j}=1\quad(j\in V).
\]
Set $p_j^{(0)}=1$ for all $j\in V$. For $t=1,\ldots,|I|$, compute
\[
 p_j^{(t)}=\max\left\{p_j^{(t-1)},
              \ \max_{i\in V}p_i^{(t-1)}\rho_{ij}\right\}
       \qquad(j\in V).
\]
Writing $p_j=p_j^{(|I|)}$, return $a$, the prices $\pi_i=p_i/p_*$,
and the valuations $w_i(h)=\pi_i v_i(h)/v_i(a_i)$.
For correctness, a simple cycle of real vertices cannot have product
above one by optimality of the matching. For a simple cycle through
$*$ of length at least two, use a dummy matched to a good attaining
the maximum on the arc entering $*$. Its outgoing edge still has
weight one. The cycle again cannot improve the padded matching, and
the loop at $*$ has weight one. Thus deleting cycles cannot reduce a
walk's weight, even when an edge has weight zero. The recurrence
gives the maximum weight of a walk with at most $t$ edges, starting
anywhere and ending at $j$. A simple walk has at most $|I|$ edges,
so the last row gives the unrestricted maxima. Appending one edge
shows $p_i\rho_{ij}\le p_j$. In particular $p_*\le p_j$, and the three
inequalities in~\eqref{eq:prices} follow.

\subsection{Constructing the initial partial allocation}

Algorithm~\ref{alg:initial-passes} implements the two passes of
Lemma~\ref{lem:two-passes}. Algorithm~\ref{alg:price-pass} implements
the exchanges of Lemma~\ref{lem:initial}. In the
second subroutine, an expensive trim for agent $i$ is one whose
value to $i$ exceeds $rD_i$. By the proof of the latter lemma,
such a trim necessarily belongs to a second-pass triple and
contains its current matched good. This makes the choice of
$\ell$ in the algorithm well-defined whenever it is required.

\begin{algorithm}[htbp]
\SetAlgoLined
\DontPrintSemicolon
\KwIn{The core $I,G$, matched goods $a$, prices $\pi$, and rescaled values $w$.}
\KwOut{A partial allocation $B$.}
$B_i\gets\{a_i\}$ for each $i\in I$\;
\ForEach{$i\in I$ with $\pi_i\le r$ and $P(B)\ne\emptyset$}{
  $g\gets$ a favourite good of $i$ in $P(B)$\;
  \If{$\max\{b,\pi_i\}+w_i(g)>r\pi_i$}{$B_i\gets\{a_i,g\}$\;}
}
\ForEach{$j\in I$ with $\pi_j>r$ and $|P(B)|\ge2$}{
  $g,h\gets$ the two favourite distinct goods of $j$ in $P(B)$, best first\;
  \If{$w_j(g)>(r-1)\pi_j$ \textbf{and} $w_j(g)+w_j(h)\ge d_0$}{
    \eIf{some $i$ has $B_i=\{a_i\}$, $\pi_i<b$ and $w_i(a_j)\ge b$}{
      $(B_i,B_j)\gets(\{a_j\},\{a_i,g,h\})$ for such an $i$\;
    }{
      $B_j\gets\{a_j,g\}$\;
    }
  }
}
\Return $B$\;
\caption{\textsc{InitialPasses} forms pairs and triples from the matching.}
\label{alg:initial-passes}
\end{algorithm}
\FloatBarrier

After the first two passes, an agent $i$ has an exception if a good
$g_i$ in the pool violates the insertion condition~\eqref{eq:addition}.
Lemma~\ref{lem:two-passes} shows that each agent has at most one such good.
Different agents can have the same exception.

\begin{algorithm}[htbp]
\SetAlgoLined
\DontPrintSemicolon
\KwIn{The partial allocation from \textsc{InitialPasses}, and $a,\pi,w$.}
\KwOut{A partial allocation $B$, its watchers $W$, and their exceptional goods.}
$J\gets$ the agents with an exception in the current pool, recording $g_i$ for each $i\in J$\;
\ForEach{$i\in J$ in nonincreasing order of $\pi_i$, with $g_i\in P(B)$}{
  $D_i\gets(1+\pi_i/r)/r$, \quad $z_i\gets w_i(g_i)$\;
  $j\gets$ the owner of an expensive trim, if one exists\;
  $h\gets$ a favourite of $i$ in $P(B)\setminus\{g_i\}$, if nonempty\;
  \If{$j$ exists \textbf{or} ($h$ exists \textbf{and} $z_i+w_i(h)\ge D_i$)}{
    \eIf{$w_k(a_i)\le ru_k$ for all $k$}{
      $B_i\gets\{a_i,g_i\}$\;
    }{
      Choose $k$ with $w_k(a_i)>ru_k$\;
      \eIf{$j$ exists}{
        Let $\ell$ be the matched good of $B_j=\{\ell,e,f\}$\;
        $(B_k,B_i,B_j)\gets(\{a_i\},\{\ell,g_i\},\{a_k,e,f\})$\;
      }{
        $(B_k,B_i)\gets(\{a_i\},\{a_k,g_i,h\})$\;
      }
    }
  }
}
$W\gets\{i\in J:g_i\in P(B)\}$\;
\Return $(B,W,(g_i)_{i\in W})$\;
\caption{\textsc{ResolveExceptions} visits the exceptional agents in price order.}
\label{alg:price-pass}
\end{algorithm}
\FloatBarrier

\subsection{Completing the allocation}

Algorithm~\ref{alg:completion} implements the three operations
in the proof of Lemma~\ref{lem:complete}. Its state consists of the
current partial allocation and the set $W$ of agents still
tracking an exceptional good. The goods $g_i$ and the numerical
parameters associated with the initial watchers remain fixed;
only the current bundles and membership of $W$ change.

At initialization, for each watcher set
\[
 z_i=w_i(g_i),\qquad
 s_i=(1+\pi_i/r)/r-z_i,\qquad
 d_i=\max\{1+\pi_i/r,q+s_i\}/r .
\]
The current pool is always $P(B)$, including goods returned by a
path transfer. The following rule gives a direct implementation of
the replacement used in such a transfer.

\paragraph{Trimming a replacement bundle.}
$\textsc{TrimReplacement}(j,g,C,t)$ uses the current bundles $B$,
the original matches $a$, and the rescaled valuations $w$ from the
calling procedure. Here $j$ is the starting agent, $g$ the chosen
pool good, $C$ the source component, and $t$ its receiving thresholds.
Start with $S=B_j\cup\{g\}$.
In any fixed order, visit each $h\in B_j\setminus\{a_j\}$ once.
Delete $h$ from $S$ exactly when some $i\in C$ still has
$w_i(S\setminus\{h\})>t_i$. Return the final $S$ and any
$k\in C$ with $w_k(S)>t_k$. Such a recipient exists initially
because $j$ has an incoming edge in its nonsingleton component.
Each deletion preserves the existence of a recipient. If $h$ was
retained, removing it failed the test for every member of $C$ at
the time of its visit. The set only shrinks afterwards, so
non-negativity makes the same test fail at the end. Thus the goods
retained in addition to $a_j,g$ form an inclusion-minimal set for
acceptance by \emph{any} member of $C$ as required in
Lemma~\ref{lem:complete}.

\begin{algorithm}[htbp]
\SetAlgoLined
\DontPrintSemicolon
\KwIn{The output of \textsc{ResolveExceptions}, and $a,\pi,w$.}
\KwOut{A complete allocation $B$ on the core.}
Compute $d_i$ for each initial watcher as above\;
\While{$P(B)\ne\emptyset$}{
  Choose $g\in P(B)$ and put $W_g\gets\{i\in W:g_i=g\}$\;
  Set $t_i\gets\max\{b,u_i\}$ for $i\notin W$, $t_i\gets q$ for $i\in W\setminus W_g$, and $t_i\gets d_i$ for $i\in W_g$\;
  Form the directed graph with arcs $i\to j$ for $i\ne j$ and $w_i(B_j)>t_i$\;
  Choose a source strongly connected component $C$\;
  \uIf{$C=\{k\}$}{
    $B_k\gets B_k\cup\{g\}$\;
    \If{$k\in W$ and $w_k(B_k)\ge q$}{$W\gets W\setminus\{k\}$\;}
  }
  \uElseIf{$C\cap W_g=\emptyset$}{
    Choose a directed cycle in $C$ and give each participant their successor's old bundle\;
    Remove the cycle participants from $W$\;
  }
  \Else{
    Choose $j_0\in C$ with $u_{j_0}<b$ if possible, and otherwise choose $j_0\in C\cap W_g$\;
    $(S,k)\gets\textsc{TrimReplacement}(j_0,g,C,t)$\;
    Find a simple directed path $j_0,\ldots,j_\ell=k$ in $C$\;
    Simultaneously give $B_{j_{h+1}}$ to $j_h$ for $0\le h<\ell$ and give $S$ to $k$\;
    $W\gets W\setminus(\{j_0,\ldots,j_\ell\}\cup W_g)$\;
  }
}
\Return $B$\;
\caption{\textsc{CompleteAllocation} assigns the goods remaining after the three passes.}
\label{alg:completion}
\end{algorithm}

\FloatBarrier
\section{Runtime analysis}\label{sec:runtime}

\begin{theorem}[Running time of \textsc{Allocate}]\label{thm:runtime}
Algorithm~\ref{alg:allocation} computes a complete $\alpha$-EFX allocation
using $O(n^3m^2)$ arithmetic operations, comparisons, and
operations on agent and good indices. For rational valuations,
the algorithm runs in strongly polynomial time.
\end{theorem}

\begin{proof}
We count arithmetic operations, comparisons, and operations on agent
and good indices. For real valuations we use exact arithmetic with
the fixed constant $r=1/\alpha$. The input is the matrix of
singleton values. Here $n$ and $m$ denote the numbers of agents and goods in the original
input and bound the corresponding sizes of the residual instance.

\paragraph{Support and prices.}
Forming the positive-support graph takes $O(nm)$ operations.
The graph has $O(m)$ vertices and at most $nm$ edges, so
Hopcroft--Karp finds a maximum-cardinality matching in
$O(nm\sqrt{m})$ time \citep{hopcroft1973matching}.
One graph search then finds the sets in Lemma~\ref{lem:positive-core}.

For the maximum-product matching in \textsc{PriceMatching}, the
following implementation of successive augmentation works with
positive values and their reciprocals. Start with the empty matching.
For a current matching $F_k$ of size $k$, direct each unused positive
edge $i\to h$ with gain $v_i(h)$ and each matched edge $h\to i$
with gain $1/v_i(h)$. The gain of a walk is the product of its gains.
Augment along a maximum-gain path from an unmatched agent to an
unmatched good, stopping when every agent is matched. Such a path
exists at each required size because the core admits a saturating
matching.

Inductively $F_k$ maximizes the product at size $k$, with the empty
product equal to one. Every directed simple cycle has gain at most
one, since flipping its edges retains size $k$. Thus a maximum-gain
augmenting walk can be made simple by removing repeated vertices.
For optimality after augmentation, take an optimal size-$(k+1)$
matching $Y$. Some path component $Q$ of $F_k\mathbin{\triangle}Y$
contains one more edge of $Y$ than of $F_k$. It is an augmenting
path for $F_k$. Flipping just $Q$ in $Y$ gives a size-$k$ matching
of product $\operatorname{prod}(Y)/\operatorname{gain}(Q)$, at most
$\operatorname{prod}(F_k)$. An augmenting path of maximum gain
therefore produces a matching with product at least
$\operatorname{prod}(Y)$. This proves the induction.

To find the path exactly, initialize labels to one at free agents
and zero elsewhere. In each synchronous relaxation, take the larger
of the old label and the maximum incoming label times its arc gain.
After $t$ rounds the labels give the best gains of walks of length
at most $t$ from any free agent, with zero denoting no walk.
After $|I|+|G|-1$ rounds a best free good has the optimum label.
Store the maximizing predecessor for each round, recover a walk, and
erase repeated vertices. This is the multiplicative form of
Bellman--Ford relaxation \citep{bellman1958routing}. Each augmentation
takes $O(nm^2)$ operations and there are at most $n$ augmentations.
For the prices, forming the arc weights takes $O(nm)$ operations.
The $|I|$ rounds on $|I|+1$ vertices take $O(n^3)$ further
operations. The whole matching stage takes $O(n^2m^2)$.

\paragraph{Initial allocation.}
Each of the three passes visits each of its agents once. Finding
two favourites takes one scan of the pool. Finding an observer or
a trim that exceeds the visitor's bound takes $O(n)$ time, since
every current bundle has at most three goods. Identifying all initial
exceptions takes $O(nm)$ time, and sorting the visitors by price takes
$O(n\log(n+1))$. Thus Algorithms~\ref{alg:initial-passes}
and~\ref{alg:price-pass} together take $O(nm)$ operations.

\paragraph{Number of completion iterations.}
We bound the number of path reassignments, direct insertions,
and cycle rotations separately. Let $T_{\mathrm{path}}$,
$T_{\mathrm{ins}}$, and $T_{\mathrm{cyc}}$ denote the numbers of
these three kinds of operation, respectively. The argument below
applies to any finite prefix of the completion procedure.

Recall that
\[
    A=\{i\notin W:u_i\geq b\}.
\]
No operation removes an agent from $A$, and every path
reassignment adds its starting agent to $A$. Consequently,
\[
    T_{\mathrm{path}}\leq n.
\]
A direct insertion reduces the size of the pool by one, and a
cycle rotation does not change the pool. A path reassignment
can return at most $m$ goods to the pool. Since the initial
pool has at most $m$ goods and the size of the pool is always
non-negative, we obtain
\[
    T_{\mathrm{ins}}
    \leq m+mT_{\mathrm{path}}
    \leq m(n+1)
    =O(nm).
\]

It remains to bound the number of cycle rotations. For a partial
allocation $B$, define
\[
    \Phi(B)=\sum_i
    \bigl|\{j:w_i(B_j)>w_i(B_i)\}\bigr|,
\]
where both indices range over the agents of the residual
instance. Thus $\Phi(B)$ is the total number of ordered pairs
of agents in which the first agent strictly prefers the second
agent's bundle to their own. In particular,
\[
    0\leq\Phi(B)\leq n(n-1).
\]

A cycle rotation permutes the current bundles and strictly
increases the utility of every participant. A nonparticipant
keeps the same bundle, and hence their contribution to
$\Phi(B)$ does not change. For a participant, the number of
bundles worth strictly more than their own cannot increase,
and the bundle they receive was counted before the rotation
but is no longer counted afterwards. Their contribution thus
strictly decreases. It follows that every cycle rotation
reduces $\Phi(B)$ by at least one.

Consider next a direct insertion, in which a good is added to
agent $k$'s bundle. For every agent $i\ne k$, the utility of $i$
and all bundles other than $B_k$ remain unchanged. The only
bundle that can newly contribute to their count is $B_k$.
Thus their contribution increases by at most one. Agent $k$'s
own utility cannot decrease, and every other bundle remains
unchanged, so their contribution cannot increase. Consequently,
a direct insertion increases $\Phi(B)$ by at most $n-1$.

A path reassignment may increase the potential, but since
$\Phi(B)$ always lies between zero and $n(n-1)$, any single
path reassignment increases it by at most $n(n-1)$.
Combining these estimates and using the non-negativity of the
final potential gives
\begin{align*}
    T_{\mathrm{cyc}}
    &\leq n(n-1)+(n-1)T_{\mathrm{ins}}
                    +n(n-1)T_{\mathrm{path}}\\
    &=O(n^2m+n^3)\\
    &=O(n^2m),
\end{align*}
where the last equality uses the standing assumption $m>n$.

The total number of completion iterations is therefore
\[
    T_{\mathrm{path}}+T_{\mathrm{ins}}+T_{\mathrm{cyc}}
    =O(n^2m).
\]
Since this bound holds for any finite prefix, it also gives a
quantitative proof of termination. The completion procedure
maintains admissibility and has an available operation whenever
the pool is nonempty. It must therefore terminate with an
empty pool. The correctness results proved above give the
required complete $\alpha$-EFX allocation on the residual
instance, which the positive matching reduction extends to the
original instance.

\paragraph{Cost of one iteration.}
Store the goods of each current bundle and the pool. Initialize an
$|I|$-by-$|I|$ table to zero and, for each assigned good, add its
value for each observer to the column of its owner. This computes
all $w_i(B_j)$ in $O(nm)$ operations. Form the threshold graph
from the table. Its strongly connected components take $O(n^2)$ time
by Tarjan's algorithm \citep{tarjan1972depth}. Scanning the arcs
finds a source component, and one further graph search gives the
required cycle or path. During \textsc{TrimReplacement}, keep the
current sums $w_i(S)$ for $i\in C$. Testing a deletion and updating
all sums when the test succeeds cost $O(n)$ per examined good,
hence $O(nm)$ in total. The simultaneous reassignments and pool
updates can be carried out by one scan of the bundles and their
goods. Each iteration therefore costs $O(nm)$ operations. Together
with the bound of $O(n^2m)$ on the number of iterations, this
gives a total cost of $O(n^3m^2)$ for completion. The preceding
matching and initial-allocation stages take $O(n^2m^2)$ and
$O(nm)$ operations, respectively. Hence the overall number of
operations is $O(n^3m^2)$.

\paragraph{Exact arithmetic and encoding length.}
For rational input, first multiply each row by the product of its
denominators. This takes $O(nm)$ operations and gives integral
values of polynomial bit length. Positive row scaling preserves
all fairness comparisons. Both the scaled and original products
of a matching saturating $I$ differ by the same positive factor.
The augmenting-path labels are products of at most $n+m$ input
entries or reciprocals, recomputed from the current arcs at each
augmentation. A price label uses at most $n$ ratios of entries.
All numerators and denominators in these computations therefore
have polynomial length, even as unreduced fractions. For each
observer, store a normalized bundle value as its fixed rational row
scale times the integer sum of the goods in that bundle. The sums
used in a deletion are sums over subsets of the goods and have
polynomial length as well.

The constants $q,b,d_0$ and the threshold calculations involve the
fixed field $\mathbb Q(r)$. In the basis $1,r,r^2$, its arithmetic
uses a constant number of rational operations by the cubic relation
$r^3-2r^2+2r-2=0$. Comparisons can also be exact in a constant
number of rational operations. The cubic has no rational root
among $\pm1,\pm2$, and its other two roots form a nonreal conjugate
pair by Section~\ref{sec:parameters}. For a nonzero field element
$x$, the determinant of its $3$-by-$3$ multiplication matrix is the
product of its three conjugates. Indeed, evaluation at the three
distinct roots diagonalizes multiplication over $\mathbb C$.
The product is the real value of $x$ times a positive squared
modulus, so the determinant gives its sign and also detects equality
at zero. After normalization, the algorithm forms its thresholds and
comparisons from the indicated row sums, prices, and fixed constants
by a bounded number of field operations per expression. It never
iterates field multiplications on a changing threshold. Consequently
all rational coordinates and determinants have polynomial bit
length, and the stated operation count applies.
\end{proof}


\end{document}